\documentclass[12pt]{iopart}
\usepackage{iopams}
\usepackage{color}

\usepackage[mathscr]{eucal}

\newtheorem{theorem}{Theorem}
\newtheorem{lemma}[theorem]{Lemma}

\newtheorem{proposition}[theorem]{Proposition}

\newtheorem{definition}[theorem]{Definition}
\newtheorem{example}[theorem]{Example}

\newtheorem{remark}[theorem]{Remark}

\newcommand{\Z}{{\mathbb Z}}

\begin{document}
\title{Fermionic partition functions for a periodic soliton cellular automaton}
\author{Atsuo Kuniba}
\address{Institute of Physics, University of Tokyo, Tokyo 153-8902, Japan}
\ead{atsuo@gokutan.c.u-tokyo.ac.jp}

\author{Taichiro Takagi}
\address{Department of Applied Physics, National Defense Academy, Kanagawa 239-8686, Japan}
\ead{takagi@nda.ac.jp}
\begin{abstract}
Fermionic formulas in combinatorial Bethe ansatz consist of 
sums of products of $q$-binomial coefficients.
There exist refinements without a sum that are known to yield  
partition functions of box-ball systems with a prescribed soliton content.
In this paper, such a refined fermionic formula is extended to the 
periodic box-ball system and a $q$-analogue 
of the Bethe root counting formula
for XXZ chain at $\Delta=\infty$.
\end{abstract}
\maketitle

\section{Introduction}\label{sec:1}
\subsection{Main result}\label{ss:mr} 
Combinatorial aspects of integrable systems have attracted 
considerable interest, constituting the field sometimes called 
``physical combinatorics" (see e.g.  \cite{PC,PC2}).
In this paper we establish a new {\em fermionic formula} 
(Theorem \ref{th:main}):
\begin{equation}\label{eq:main}
\sum_{p \in P^\circ_{L,\lambda}} 
q^{E_{\rm path}(p)} =
q^{\psi (\lambda)} \frac{[L]}{[p_1]}
\prod_{i=1}^{s}
\left[
\begin{array}{c}
p_i+m_i-1\\
m_i
\end{array}
\right],
\end{equation}
where $q$ is an indeterminate,
$L$ is a positive integer and $\lambda$ is a Young diagram
such that $|\lambda | \le L/2$.
The numbers $m_i$ and $s$ are determined from $\lambda$ 
by Figure \ref{fig:1}, from which $p_i$ (\ref{eq:may28_2}) and 
$\psi(\lambda)$ (\ref{eq:may28_3}) are also specified.
The bracket symbols in the right hand side are 
defined in (\ref{qb}).

The left hand side of (\ref{eq:main}) 
is a partition function of the 
length $L$ {\em periodic box-ball system} (section \ref{subsec:3_1})
with a prescribed {\em soliton content} $\lambda$.
A state $p$ of the box-ball system, 
we call it a {\em path} in this paper, 
is just an array of $1$ and $2$. 
The paths are subject to commuting time evolutions $T_1, T_2, \ldots$.
See Example \ref{ex:2}.
Regarding 1's as background,  one observes that there are 
solitons (consecutive 2's) with amplitudes $4,3,2,1$ 
moving to the right under the periodic boundary condition.
They get reshuffled locally and temporarily under the collisions 
but regain their original amplitudes after the events.
In fact the list of properly defined amplitudes are known to be conserved
and that defines 
the soliton content\footnote{It is $\lambda =(4321)$ 
in Example \ref{ex:2}.}.
The $P^\circ_{L,\lambda}$ in (\ref{eq:main}) 
denotes the set of length $L$ paths having the soliton content $\lambda$.
Finally $E_{\rm path}(p)$, the {\em energy} of a path $p$, is 
given in Definition \ref{def:Oct8_1}.

\subsection{Background and related results}\label{ss:rr}
As a guide to the formula (\ref{eq:main}),  
we recall some background and related results obtained earlier.
We shall concentrate on the  $sl_2$ 
case throughout the paper.

Generating functions of energy over paths, denoted by $X$, 
are known as the one-dimensional configuration sums 
in Baxter's corner transfer matrix method \cite{Ba}.
On the other hand, the combinatorial version \cite{KKR, KR} of the 
Bethe ansatz \cite{Be} provides the same quantity with another counting 
based on string hypothesis.
It leads to the so called fermionic form $\EuScript{M}$.
Then the physically suggested identity $X=\EuScript{M}$ can indeed 
be proved by the Kerov-Kirillov-Reshetikhin (KKR) bijection 
\cite{KKR, KR} between paths and  
{\em rigged configurations} which are combinatorial objects
labeling the monomials in $\EuScript{M}$.
See for example the reviews by M. Okado and A. Schilling in \cite{PC2}
and references therein.

The fermionic form is actually a sum 
$\EuScript{M}=\sum_\lambda \EuScript{M}(\lambda)$,
where each summand is a product of 
$q$-binomial coefficients reflecting the fermionic nature of Bethe roots.
The sum gathers the contributions 
from all possible decompositions of the Bethe roots into strings.  
The $\lambda$ in the sum 
is a Young diagram called {\em configuration}
labeling such string contents.
A natural problem is then to refine the identity 
$X=\EuScript{M}$ to $X(\lambda)=\EuScript{M}(\lambda)$.
By the construction, the paths responsible for $X(\lambda)$
are the restriction of the image of the KKR map 
to those rigged configurations having configuration $\lambda$.
However is there any, hopefully ``good physical" meaning of 
these paths ?

This is a point where one can develop a further insight into 
fermionic formulas 
in the light of the soliton cellular automaton 
known as the box-ball system \cite{TS}.
According to such studies initiated in \cite{KOTY} and accomplished in 
Theorem \ref{th:t} for $sl_2$ case \cite{Takagi1},
those paths are characterized by the conserved quantities 
or equivalently by the soliton content $\lambda$.
The $X(\lambda)$ thereby acquires the meaning of the partition function 
of the {\em level set} of the box-ball system. 
These features may be schematically stated as 
``solitons = strings", where 
the both sides can be defined mathematically 
within the framework of the box-ball system
and the rigged configuration, respectively.

\subsection{Comparison with earlier results and 
layout of the paper}\label{ss:xx}
The story (Theorem \ref{th:t}) in the previous subsection 
concerns the string hypothesis in the XXX Heisenberg spin chain.
The relevant box-ball system is 
the one on the infinite lattice \cite{TS} 
truncated to the finite segment 
$[1,L]=\{j\in \Z\mid 1 \le j \le L\}$.

Our formula (\ref{eq:main}) provides an extension 
of such results in which 
XXX and the box-ball system are replaced with 
XXZ and the periodic box-ball system \cite{YYT,KTT}, respectively.
More specifically,
the right hand side 
of (\ref{eq:main}) with $q=1$ 
precisely reproduces the counting formula of 
the Bethe roots for the XXZ Hamiltonian 
$\sum_{j\in \Z_L}(\sigma^x_j\sigma^x_{j+1}+
\sigma^y_j\sigma^y_{j+1}+
\Delta \sigma^z_j\sigma^z_{j+1})$ at $\Delta = \infty$ 
\cite[eq. (3.11)]{KN}.
With regard to the periodic box-ball system, 
it is defined on the ring $\Z_L$, and 
the level set $P^\circ_{L,\lambda}$ is equivalent to 
that introduced in \cite{KTT}, where the 
first complete solution of the 
initial value problem was obtained.
A bold summary of these aspects 
is given in the following table\footnote{
A conceptual explanation of the correspondence 
between these values of $\Delta$ and the
boundary conditions of the box-ball systems is 
yet to be found. }.

\begin{center}
\begin{tabular}{c|c|c}
string hypothesis & box-ball system &  fermionic form  \\
\hline 
&&\vspace{-0.4cm}\\
XXX \;$(\Delta=1)$ & $[1,L]$  & 
$\EuScript{M}(\lambda, L), \,
\EuScript{M}^{+}(\lambda, L)$ \; eq. (\ref{mm})\\
&& \vspace{-0.4cm}\\
\hline
&& \vspace{-0.4cm}\\
XXZ $(\Delta=\infty)$ & $\Z_L$  & 
$\EuScript{M}^{\circ}(\lambda, L)$ \;\; eq. (\ref{eq:nov1_2})
\end{tabular}
\end{center}

Compared with the fermionic forms
$\EuScript{M}(\lambda, L)$ and 
$\EuScript{M}^{+}(\lambda, L)$,
the new one $\EuScript{M}^{\circ}(\lambda, L)$
(= right hand side of eq. (\ref{eq:main})) takes the 
contributions of the solitons wrapping 
around the spurious boundary of the segment $[1,L]$.
Compare Examples \ref{ex:oct9_3} and \ref{ex:oct19_1}.
Coping with them systematically 
and factorizing the whole level set partition function into a
product of $q$-binomial coefficients 
require some technical analysis.
It will be done in sections \ref{sec:4} and \ref{sec:5}.
Section \ref{sec:2} recalls the previous result 
(second line of the table) for comparison and 
convenience of the readers.
In section \ref{sec:3}
we describe the basic feature of the periodic box-ball system
and formulate the main result of the paper in 
Theorem \ref{th:main}.

\section{Fermionic expressions for one-dimensional 
configuration sums over non-periodic paths}\label{sec:2}
\begin{figure}[h]
\unitlength 0.1in
\begin{picture}( 26.2000, 23.0000)( 0.8000,-31.0000)
%
\special{pn 8}%
\special{pa 1600 800}%
\special{pa 1600 3000}%
\special{pa 2400 3000}%
\special{pa 2400 2600}%
\special{pa 2800 2600}%
\special{pa 2800 2000}%
\special{pa 3400 2000}%
\special{pa 3400 1200}%
\special{pa 3800 1200}%
\special{pa 3800 800}%
\special{pa 3800 800}%
\special{pa 3800 800}%
\special{pa 1600 800}%
\special{fp}%
%
\special{pn 8}%
\special{pa 2000 3090}%
\special{pa 2400 3090}%
\special{fp}%
\special{sh 1}%
\special{pa 2400 3090}%
\special{pa 2334 3070}%
\special{pa 2348 3090}%
\special{pa 2334 3110}%
\special{pa 2400 3090}%
\special{fp}%
\put(18.7000,-32.7000){\makebox(0,0)[lb]{$\ell_s$}}%
%
\special{pn 8}%
\special{pa 2000 3090}%
\special{pa 1600 3090}%
\special{fp}%
\special{sh 1}%
\special{pa 1600 3090}%
\special{pa 1668 3110}%
\special{pa 1654 3090}%
\special{pa 1668 3070}%
\special{pa 1600 3090}%
\special{fp}%
%
\special{pn 8}%
\special{pa 2610 2700}%
\special{pa 2810 2700}%
\special{fp}%
\special{sh 1}%
\special{pa 2810 2700}%
\special{pa 2744 2680}%
\special{pa 2758 2700}%
\special{pa 2744 2720}%
\special{pa 2810 2700}%
\special{fp}%
%
\special{pn 8}%
\special{pa 2620 2700}%
\special{pa 2420 2700}%
\special{fp}%
\special{sh 1}%
\special{pa 2420 2700}%
\special{pa 2488 2720}%
\special{pa 2474 2700}%
\special{pa 2488 2680}%
\special{pa 2420 2700}%
\special{fp}%
%
\special{pn 8}%
\special{pa 3600 1320}%
\special{pa 3800 1320}%
\special{fp}%
\special{sh 1}%
\special{pa 3800 1320}%
\special{pa 3734 1300}%
\special{pa 3748 1320}%
\special{pa 3734 1340}%
\special{pa 3800 1320}%
\special{fp}%
\special{pa 3600 1320}%
\special{pa 3400 1320}%
\special{fp}%
\special{sh 1}%
\special{pa 3400 1320}%
\special{pa 3468 1340}%
\special{pa 3454 1320}%
\special{pa 3468 1300}%
\special{pa 3400 1320}%
\special{fp}%
%
\special{pn 8}%
\special{pa 2310 2800}%
\special{pa 2310 3000}%
\special{fp}%
\special{sh 1}%
\special{pa 2310 3000}%
\special{pa 2330 2934}%
\special{pa 2310 2948}%
\special{pa 2290 2934}%
\special{pa 2310 3000}%
\special{fp}%
\special{pa 2310 2820}%
\special{pa 2310 2600}%
\special{fp}%
\special{sh 1}%
\special{pa 2310 2600}%
\special{pa 2290 2668}%
\special{pa 2310 2654}%
\special{pa 2330 2668}%
\special{pa 2310 2600}%
\special{fp}%
%
\special{pn 8}%
\special{pa 2720 2400}%
\special{pa 2720 2600}%
\special{fp}%
\special{sh 1}%
\special{pa 2720 2600}%
\special{pa 2740 2534}%
\special{pa 2720 2548}%
\special{pa 2700 2534}%
\special{pa 2720 2600}%
\special{fp}%
\special{pa 2720 2400}%
\special{pa 2720 2000}%
\special{fp}%
\special{sh 1}%
\special{pa 2720 2000}%
\special{pa 2700 2068}%
\special{pa 2720 2054}%
\special{pa 2740 2068}%
\special{pa 2720 2000}%
\special{fp}%
\special{pa 3720 1000}%
\special{pa 3720 1200}%
\special{fp}%
\special{sh 1}%
\special{pa 3720 1200}%
\special{pa 3740 1134}%
\special{pa 3720 1148}%
\special{pa 3700 1134}%
\special{pa 3720 1200}%
\special{fp}%
\special{pa 3720 1000}%
\special{pa 3720 800}%
\special{fp}%
\special{sh 1}%
\special{pa 3720 800}%
\special{pa 3700 868}%
\special{pa 3720 854}%
\special{pa 3740 868}%
\special{pa 3720 800}%
\special{fp}%
\put(25.000,-29.3000){\makebox(0,0)[lb]{$\ell_{s-1}-\ell_s$}}%
\put(34.7000,-15.4000){\makebox(0,0)[lb]{$\ell_1-\ell_2$}}%
\put(20.1000,-28.9000){\makebox(0,0)[lb]{$m_s$}}%
\put(23.0000,-24.1000){\makebox(0,0)[lb]{$m_{s-1}$}}%
\put(35.0000,-11.0000){\makebox(0,0)[lb]{$m_1$}}%
\put(11.8000,-19.6000){\makebox(0,0)[lb]{$\lambda = $}}%
\end{picture}%
\caption{The Young diagram.}
\label{fig:1}
\end{figure}
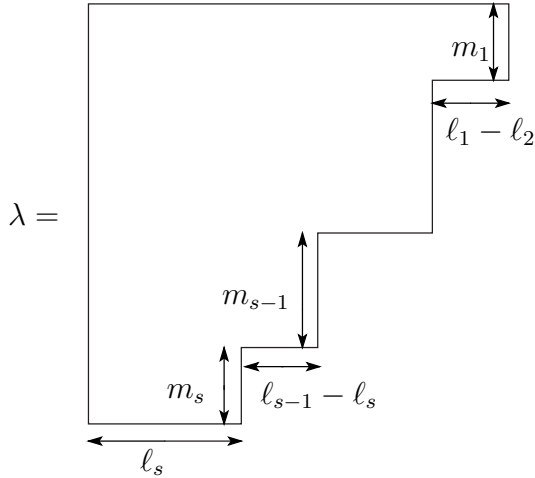

Fix any Young diagram $\lambda$.
Suppose it has $m_i$ rows of length $\ell_i$ for $1 \leq i \leq s$, and $\ell_1 > \ldots > \ell_s$.
See Figure \ref{fig:1}.
We define
\begin{eqnarray}
p_i(\lambda) :&=& L - 2 \sum_{j = 1}^s \min(\ell_i,\ell_j) m_j, 
\label{eq:may28_2}\\
%
\psi (\lambda): &=& \sum_{i,j = 1}^s \min(\ell_i,\ell_j) m_i m_j
- \sum_{i=1}^s  \ell_i m_i, \label{eq:may28_3}
\end{eqnarray}
which depend on $\lambda$ and a fixed integer $L$ satisfying $L \geq 2 \sum_{j = 1}^s \ell_j m_j$.
The $p_i = p_i(\lambda)$'s are called vacancy numbers.

A Young diagram $\lambda$ is also regarded as a partition $\lambda = (\lambda_1 \lambda_2 \ldots)$ where $\lambda_i$ is 
called a {\em part} and equals to the length of the $i$-th row of $\lambda$.
Parts are obeying the condition $\lambda_1 \geq \lambda_2 \geq \ldots \geq 0$ and those parts satisfying $\lambda_i = 0$ are not explicitly written.
If $|\lambda| := \sum_i \lambda_i = M$ then $\lambda$ is call a partition of $M$ and denoted by $\lambda \vdash M$.
We say that there are $\lambda_i$ {\em boxes} in the $i$-th row of the Young diagram $\lambda$.

By using this terminology, the $\psi (\lambda)$ is also calculated as follows.
Fill $0$'s, $2$'s, $4$'s ... into the boxes of the first, second, third, ... rows of 
the Young diagram $\lambda$.
Then the sum of these integers gives $\psi (\lambda)$.
\begin{example}\label{ex:nov2_1}
There are three partitions of $3$, which are $(3), (111)$ and $(21)$.
For these partitions, $\psi ((3)) = 0+0+0=0, \psi ((111)) = 0+2+4=6$ and $\psi ((21)) = 0+0+2=2$. 
\end{example}

Let
$\boldsymbol{x}_i = ( x_{i,j} )_{1 \leq j \leq m_i}, \, x_{i,j} \in \Z$
be a set of integers and set
$\boldsymbol{x} = (\boldsymbol{x}_i)_{1 \leq i \leq s}$.
The integers $x_{i,j}$ are called quantum numbers\footnote{Conventionally
the quantum numbers are given by $x_{i,j} - \ell_i$.} or riggings.
We set
\begin{eqnarray}
\Omega (\lambda, L) &=& \{ \boldsymbol{x} \in \Z^{\mathcal N} |
0 \leq x_{i,1} \leq \cdots \leq x_{i,m_i} \leq
\ell_i + p_i \quad \mbox{for} \quad 1 \leq i \leq s \}, \nonumber \\
\label{eq:may27_4} \\
\Omega^+ (\lambda, L) &=& \{ \boldsymbol{x} \in \Z^{\mathcal N} |
 \ell_i \leq x_{i,1} \leq \cdots \leq x_{i,m_i} \leq
\ell_i + p_i \quad \mbox{for} \quad 1 \leq i \leq s \}, \nonumber \\
\label{eq:may27_5}
\end{eqnarray}
where ${\mathcal N} = m_1 + \cdots + m_s$.

\begin{definition}[Energy of rigged configuration]
\label{def:may12_4}
For any pair $(\lambda, \boldsymbol{x})$
the quantity
$E_{\rm RC}(\lambda, \boldsymbol{x}):=
\psi (\lambda) + 
\sum_{i=1}^s \sum_{j=1}^{m_i} x_{i,j}$ is called
its {\em energy}.
\end{definition}

Let
\begin{eqnarray}
\EuScript{M}(\lambda, L) &:=& \sum_{\boldsymbol{x} \in \Omega (\lambda, L)} 
q^{E_{\rm RC}(\lambda,\boldsymbol{x})},\\
\EuScript{M}^+(\lambda, L) &:=& \sum_{\boldsymbol{x} \in \Omega^+ (\lambda, L)} 
q^{E_{\rm RC}(\lambda,\boldsymbol{x})}.
\end{eqnarray}

For any integers $n,m (n \geq m \geq 0)$ 
we use the notation:
\begin{equation}\label{qb}
[n] = \frac{1-q^n}{1-q},\quad
\left[
\begin{array}{c}
n\\
m
\end{array}
\right] = \frac{(q)_n}{(q)_m(q)_{n-m}} \;\;\; 
(q\!-\!{\rm binomial \, coefficient}),
\end{equation}
where $(q)_n = (1-q) (1-q^2)\cdots(1-q^{n})$.
It is easy to see
${ p+m \brack m } = \sum_{\mu} q^{|\mu|}$
where the sum is taken over all partitions $\mu$ which has at most $m$ parts and the largest part $\leq p$ \cite{A, St}.
Hence we have:
\begin{lemma}\label{lem:may28_4}
The following relations hold,
\begin{equation}\label{mm}
\EuScript{M}(\lambda, L) = 
q^{\psi(\lambda)}
\prod_{i = 1}^s { p_i + m_i + \ell_i
\brack  m_i },\quad
\EuScript{M}^+(\lambda, L) = 
q^{\psi(\lambda) + |\lambda |}
\prod_{i = 1}^s { p_i + m_i 
\brack  m_i }.
\end{equation}
\end{lemma}

Let $p = b_1 \ldots b_L$ be a sequence of letters $1$ and $2$ 
where the number of $1$'s is not less than that of $2$'s. 
Any such sequence is called a {\em path}.
We emphasize that the above condition on the numbers of the letters $1$ and $2$ are always imposed throughout this paper for any path.
If $b_1 \ldots b_i$ is a path for any $1 \leq i \leq L$, then $p = b_1 \ldots b_L$ is called a {\em highest path}.
Let $P_{L,M}$ be the set of all paths of length $L$ with $ M (\leq L/2)$ $2$'s
\begin{equation*}
P_{L,M} = \{ b_1 \ldots b_L | b_i \in \{ 1,2 \}, \# \{ i | b_i = 2 \} = M  \},
\end{equation*}
and $P^+_{L,M}$ be the set of all highest paths $\subset P_{L,M}$.

\begin{definition}[Energy of path]
\label{def:Oct8_1}
For any path $p= b_1 \ldots b_L \in P_{L,M}$
the quantity
$E_{\rm path}(p):=
\sum_{j=1}^{L-1} j \theta (b_j < b_{j+1})$ is called
its {\em energy}, where $\theta (\mbox{true}) = 1, \theta (\mbox{false}) = 0$.
\end{definition}
The following results are well known
\begin{equation}
\sum_{p \in P_{L,M}} q^{E_{\rm path}(p)} = {L \brack M}, \quad
\sum_{p \in P^+_{L,M}} q^{E_{\rm path}(p)} = {L \brack M}-{L \brack M-1}.
\end{equation}

Let $P_{L,\lambda} = \Psi^{-1}(\{ \lambda \} \times \Omega(\lambda,L))$ and
$P^+_{L,\lambda} = \Psi^{-1}(\{ \lambda \} \times \Omega^+(\lambda,L))$,
where $\Psi$ is a bijective map to be defined in section \ref{subsec:4_5}.

\begin{example}\label{ex:oct9_3}
Let $L=6,|\lambda | =3$. Then
\begin{eqnarray*}
P_{6, (3)} &=& \{ 222111, 122211, 112221, 111222 \}, \\
P_{6, (111)} &=& \{ 212121, 212112, 211212, 121212 \},\\
P_{6, (21)} &=& \{ 212211, 221211, 211221, 221121, \\
&& \, 121221, 211122, 221112, 122121, \\
&& \, 121122, 122112, 112122, 112212 \}.
\end{eqnarray*}
\end{example}
By Theorem \ref{th:oct9_1} we have
$\bigsqcup_{\lambda \vdash M} P_{L,\lambda} = P_{L,M}$ and
$\bigsqcup_{\lambda \vdash M} P^+_{L,\lambda} = P^+_{L,M}$.
We define
\begin{eqnarray}
X(\lambda, L) &:=& \sum_{p \in P_{L,\lambda}} 
q^{E_{\rm path}(p)},\\
X^+(\lambda, L) &:=& \sum_{p \in P^+_{L,\lambda}} 
q^{E_{\rm path}(p)}.
\end{eqnarray}

\begin{example}
By Example \ref{ex:oct9_3} it is easy to check
\begin{eqnarray*}
X((3),6) &=& 1 +q + q^2 + q^3, \\
X((111),6) &=& q^6 +q^7 + q^8 + q^9, \\
X((21),6) &=& q^2 +2 q^3 + 3 q^4 + 3 q^5 + 2 q^6 + q^7.
\end{eqnarray*}
\end{example}
As a result of Theorems \ref{th:oct9_1} and \ref{th:oct9_2}, we obtain:
\begin{theorem}[\cite{Takagi1}]\label{th:t}
\begin{eqnarray*}
X(\lambda, L) &=& \EuScript{M}(\lambda, L),\\
X^+(\lambda, L) &=& \EuScript{M}^+(\lambda, L).
\end{eqnarray*}
\end{theorem}

\begin{example}
By Lemma \ref{lem:may28_4} we obtain
\begin{eqnarray*}
\EuScript{M}((3),6) &=& q^0 {0+1+3 \brack 1} = 1 +q + q^2 + q^3, \\
\EuScript{M}((111),6) &=& q^6 {0+3+1 \brack 3} = q^6 +q^7 + q^8 + q^9, \\
\EuScript{M}((21),6) &=& q^2 {0+1+2 \brack 1} {2+1+1 \brack 1} = q^2 +2 q^3 + 3 q^4 + 3 q^5 + 2 q^6 + q^7.
\end{eqnarray*}
\end{example}

\section{A periodic soliton cellular automaton and its partition functions}\label{sec:3}
\subsection{The periodic box-ball system}\label{subsec:3_1}
The periodic box-ball system (pBBS) is a one-dimensional cellular automaton with periodic boundary conditions.
Denote by ${\mathcal P} = \sqcup_M P_{L,M}$ the set of all paths of length $L$.
We can define a commuting family of time evolutions $T_k \, (k=1,2,\ldots)$ acting on ${\mathcal P}$.

Let $B_k$ be the set of all one-row semistandard tableaux of length $k$ with entries $1$ and $2$.
For instance, $B_1 = \{ 1,2 \}, B_2 = \{ 11, 12, 22 \}$ and $B_3 = \{ 111, 112, 122, 222 \}$.
The combinatorial $R$ map $R: B_k \times B_1 \rightarrow B_1 \times B_k$ is defined as follows.
Depict the relation $R(x,y) = (\tilde{y}, \tilde{x})$ by 
\begin{equation*}
\begin{picture}(90,40)(-20,-9)
\unitlength 0.4mm
\put(0,10){\line(1,0){20}}\put(-6,8){$x$}\put(22,8){${\tilde x}$.}
\put(10,0){\line(0,1){20}}\put(8,25){$y$}\put(8,-9){${\tilde y}$} 
\end{picture}
\end{equation*}
Then the definition of $R$ is given by
the diagrams in Figure \ref{fig:2}.
\begin{figure}[h]
\begin{picture}(130,120)(-60,15)
\unitlength 1.1mm
\multiput(0,3)(0,22){2}{
\multiput(2,10)(65,0){2}{\line(1,0){16}}
\multiput(10,6)(65,0){2}{\line(0,1){8}}}

\put(9.3,40){$\scriptstyle 1$}
\put(-15,34){$\scriptstyle \overbrace{1 \;\cdots \cdots \; 1}^k$} 
\put(20,34){$\scriptstyle \overbrace{1 \;\cdots \cdots \; 1}^k$}
\put(9.3,28){$\scriptstyle 1$}

\put(74.3,40){$\scriptstyle 2$}
\put(50,34){$\scriptstyle \overbrace{2 \;\cdots \cdots \; 2}^k$} 
\put(85,34){$\scriptstyle \overbrace{2 \;\cdots \cdots \; 2}^k$}
\put(74.3,28){$\scriptstyle 2$}

\put(9.3,18){$\scriptstyle 1$}
\put(-18,12){$\scriptstyle \overbrace{1 \cdots 1}^{k-a}
\overbrace{2\cdots 2}^a$} 
\put(20,12){$\scriptstyle \overbrace{1 \cdots 1}^{k-a+1}
\overbrace{2\cdots 2}^{a-1}$}
\put(9.3,6){$\scriptstyle 2$}
\put(4,2){$\scriptstyle (0<a\le k)$}

\put(74.3,18){$\scriptstyle 2$}
\put(47.2,12){$\scriptstyle \overbrace{1 \cdots 1}^{k-a}
\overbrace{2\cdots 2}^a$} 
\put(85,12){$\scriptstyle \overbrace{1 \cdots 1}^{k-a-1}
\overbrace{2\cdots 2}^{a+1}$}
\put(74.3,6){$\scriptstyle 1$}
\put(69,2){$\scriptstyle (0\le a<k)$}

\end{picture}
\caption{Combinatorial $R: B_l \times B_1 \simeq B_1 \times B_l$}\label{fig:2}
\end{figure}
\vspace{3mm}
\par\noindent
By repeated use of this $R$ we define
\begin{eqnarray}
B_k \times (B_1 \times \cdots \times B_1) && \rightarrow  (B_1 \times \cdots \times B_1) \times B_k \nonumber \\
v \times (b_1 \times \cdots \times b_L) && \mapsto  (b'_1 \times \cdots \times b'_L) \times v'.\label{eq:oct9_1}
\end{eqnarray}
\begin{example}\label{ex:oct9_2}
Set $k=3, L=13, v=112$ and $b_1 \ldots b_{13} = 1122121122211$.
(We omit the symbol $\times$ here and in what follows.)
Then we have $b'_1 \ldots b'_{13} = 2111212211122$ and $v'=112$.
It is verified as
\begin{equation*}
\unitlength 1mm
{\small
\begin{picture}(130,20)(8,-11)
\multiput(0.8,0)(11.5,0){13}{\line(1,0){4}}
\multiput(2.8,-3)(11.5,0){13}{\line(0,1){6}}
  \put(1.8,5){1}
 \put(13.3,5){1}
 \put(24.8,5){2}
 \put(36.3,5){2}
 \put(47.8,5){1}
 \put(59.3,5){2}
 \put(70.8,5){1}
 \put(82.3,5){1}
 \put(93.8,5){2}
\put(105.3,5){2}
\put(116.8,5){2}
\put(128.3,5){1}
\put(139.8,5){1}

  \put(1.8,-7.5){2}
 \put(13.3,-7.5){1}
 \put(24.8,-7.5){1}
 \put(36.3,-7.5){1}
 \put(47.8,-7.5){2}
 \put(59.3,-7.5){1}
 \put(70.8,-7.5){2}
 \put(82.3,-7.5){2}
 \put(93.8,-7.5){1}
\put(105.3,-7.5){1}
\put(116.8,-7.5){1}
\put(128.3,-7.5){2}
\put(139.8,-7.5){2}

   \put(-6,-1.1){112}
  \put(5.5,-1.1){111}
   \put(17,-1.1){111}
 \put(28.5,-1.1){112}
   \put(40,-1.1){122}
 \put(51.5,-1.1){112}
   \put(63,-1.1){122}
 \put(74.5,-1.1){112}
   \put(86,-1.1){111}
 \put(97.5,-1.1){112}
  \put(109,-1.1){122}
\put(120.5,-1.1){222}
  \put(132,-1.1){122}
\put(143.5,-1.1){112.}

\end{picture}
}
\end{equation*}
\end{example}
\noindent
In this example we have $v = v'$.
In fact one can always find such $v, v' \in B_k$ with this property.
\begin{proposition}[\cite{KTT}]\label{pr:kbv}
Given any $b_1 \times \cdots \times b_L \in {\mathcal P} \subset B_1^{\times L}$,
let $v_0 \in B_k$ be the one defined in the same way as (\ref{eq:oct9_1}) by
\begin{equation*}
\overbrace{1\ldots 1}^{k} \times (b_1 \times \cdots \times b_L)  \mapsto  (b''_1 \times \cdots \times b''_L) \times v_0.
\end{equation*}
Then we have $v=v'$ in (\ref{eq:oct9_1}) when we adopt this $v_0$ as the $v$ there.
\end{proposition}
\noindent
By this choice of $v$, we define the time evolution $T_k$ as
$T_k (b_1 \cdots b_L) = b'_1 \cdots  b'_L$ by (\ref{eq:oct9_1}).
\begin{example}\label{ex:2}
We have
$T_3 (1122121122211) = 2111212211122$ by Example \ref{ex:oct9_2}.
\end{example}
Here we give an example of the time evolution of this cellular automaton. 
 
\vskip0.2cm
\noindent
\begin{center}
$t=0: \;\; 2\; 2\; 2\; 1\; 1\; 1\; 1\; 1\; 1\; 1\; 1\; 1\; 1\; 1\;2\; 2\; 2\; 2\; 1\; 1\; 1\; 2\; 1\; 1\; 2\; 2\; 1\; 1\; 1\; 1\; 1$\\
$t=1: \;\; 1\; 1\; 1\; 2\; 2\; 2\; 1\; 1\; 1\; 1\; 1\; 1\; 1\; 1\;1\; 1\; 1\; 1\; 2\; 2\; 2\; 1\; 2\; 2\; 1\; 1\; 2\; 2\; 1\; 1\; 1$\\
$t=2: \;\; 2\; 1\; 1\; 1\; 1\; 1\; 2\; 2\; 2\; 1\; 1\; 1\; 1\; 1\;1\; 1\; 1\; 1\; 1\; 1\; 1\; 2\; 1\; 1\; 2\; 2\; 1\; 1\; 2\; 2\; 2$\\
$t=3: \;\; 1\; 2\; 2\; 2\; 2\; 1\; 1\; 1\; 1\; 2\; 2\; 2\; 1\; 1\;1\; 1\; 1\; 1\; 1\; 1\; 1\; 1\; 2\; 1\; 1\; 1\; 2\; 2\; 1\; 1\; 1$\\
$t=4: \;\; 1\; 1\; 1\; 1\; 1\; 2\; 2\; 2\; 2\; 1\; 1\; 1\; 2\; 2\;2\; 1\; 1\; 1\; 1\; 1\; 1\; 1\; 1\; 2\; 1\; 1\; 1\; 1\; 2\; 2\; 1$\\
$t=5: \;\; 2\; 1\; 1\; 1\; 1\; 1\; 1\; 1\; 1\; 2\; 2\; 2\; 1\; 1\;1\; 2\; 2\; 2\; 2\; 1\; 1\; 1\; 1\; 1\; 2\; 1\; 1\; 1\; 1\; 1\; 2$\\
$t=6: \;\; 1\; 2\; 2\; 1\; 1\; 1\; 1\; 1\; 1\; 1\; 1\; 1\; 2\; 2\;2\; 1\; 1\; 1\; 1\; 2\; 2\; 2\; 2\; 1\; 1\; 2\; 1\; 1\; 1\; 1\; 1$\\
$t=7: \;\; 1\; 1\; 1\; 2\; 2\; 1\; 1\; 1\; 1\; 1\; 1\; 1\; 1\; 1\;1\; 2\; 2\; 2\; 1\; 1\; 1\; 1\; 1\; 2\; 2\; 1\; 2\; 2\; 2\; 1\; 1$\\
$t=8: \;\; 2\; 2\; 1\; 1\; 1\; 2\; 2\; 1\; 1\; 1\; 1\; 1\; 1\; 1\;1\; 1\; 1\; 1\; 2\; 2\; 2\; 1\; 1\; 1\; 1\; 2\; 1\; 1\; 1\; 2\; 2$\\
$t=9: \;\; 1\; 1\; 2\; 2\; 2\; 1\; 1\; 2\; 2\; 2\; 1\; 1\; 1\; 1\;1\; 1\; 1\; 1\; 1\; 1\; 1\; 2\; 2\; 2\; 1\; 1\; 2\; 1\; 1\; 1\; 1$
\end{center}
\vskip0.2cm
{If we denote by $p \in {\mathcal P}$ the sequence for $t=0$ then the sequence for $t=n$ stands for 
$T_4^n ( p ) \in {\mathcal P}$.}

When sufficiently separated from the other $2$'s, one can think of
a consecutive sequence of $2$'s of length $k$ as a soliton of amplitude $k$.
By an appropriate definition one can say that the number of solitons conserves for each amplitude.
In this example there are four solitons of amplitudes 1,2,3 and 4 in every time step.
In other words they have a common {\em soliton content} $\lambda = (4321)$. 

A precise definition of the soliton content of a path
$p=b_1\ldots b_L$ was presented in \cite{KTT}.
Actually there are two equivalent (but apparently different)
definitions, which we now recall quickly\footnote{The definition 
based on the ``arc rule" \cite{YYT} also gives the equivalent result.}.
The first way of determining it is to 
consider the diagram as in Example \ref{ex:oct9_2} corresponding to 
the relation (\ref{eq:oct9_1}) with $v=v'=v_0$.
($v_0$ is defined in Proposition \ref{pr:kbv}.)
Let $E_k=E_k(p)$ be the number of local vertices of the bottom right type 
in Figure \ref{fig:2} in the diagram.
Then the soliton content is the Young diagram $\lambda$ such that 
$\lambda'_1+\cdots + \lambda'_k=E_k$, where
$\lambda'_i$ denotes the length of the $i$ th column of $\lambda$
from the left. 
The second way of finding the soliton content is to convert
$p$ into a highest path $p'$ by a cyclic shift and
pick the configuration (=Young diagram $\lambda$) 
of $p'$ under the KKR bijection 
\cite{KKR,KR} from highest paths to rigged configurations. 
The result is unique even though $p'$ is not necessarily so 
for a given $p$.
It is known that the two definitions give the same  $\lambda$ 
\cite[Proposition 3.4]{KTT}.
The first and the second definitions have 
the meaning of solitons and strings respectively, 
therefore their coincidence is a manifestation of
``solitons =strings" 
in the periodic box-ball system.

Given a fixed system size $L$, the set of all paths with a common soliton content is called a {\em level set}.
We denote by ${\mathcal P}^\circ_{L,\lambda} $ the level set with soliton content $\lambda$.
Then it is known \cite{YYT,KTT} that
\begin{equation}\label{eq:nov1_1}
|{\mathcal P}^\circ_{L,\lambda}| =
\frac{L}{p_1}
\prod_{i=1}^{s}
\left(
\begin{array}{c}
p_i+m_i-1\\
m_i
\end{array}
\right).
\end{equation}
In case $p_1 = 0$, the combination $\frac{L}{p_1} {p_1+m_1-1 \choose m_1} $
is to be understood as $\frac{L}{m_1}$.
\subsection{The partition function on a level set}\label{subsec:3_2}
Let $\lambda$ be the Young diagram in Figure \ref{fig:1} and $L$ be
an integer satisfying $L \geq 2 \sum_{j = 1}^s \ell_j m_j$.
In a similar way to (\ref{eq:may27_4}) and (\ref{eq:may27_5}) we set
\begin{displaymath}
\Omega^\circ (\lambda, L) = \left\{
\boldsymbol{x} \in \Z^{\mathcal N}
\vphantom{
\begin{array}{l}
0 \leq x_{i,1} \leq \cdots \leq x_{i,m_i} \leq p_i + \min (x_{i,1},l_i-1) \\
\mbox{for} \quad 1 \leq i \leq s, \\
x_{j,m_j} - p_j \leq x_{i,1}, \mbox{for} \quad 1 \leq i < j \leq s, \\
x_{i,m_i} \leq p_i+2(\ell_i-\ell_j)+x_{j,1}-1 \mbox{for} \quad 1 \leq i < j \leq s
\end{array}
}
\right.
\left|
\begin{array}{l}
0 \leq x_{i,1} \leq \cdots \leq x_{i,m_i} \leq p_i + \min (x_{i,1},2 \ell_i-1) \\
\mbox{for} \quad 1 \leq i \leq s, \\
x_{j,m_j} - p_j \leq x_{i,1} \quad  \mbox{for} \quad 1 \leq i < j \leq s, \\
x_{i,m_i} \leq p_i+2( \ell_i-\ell_j)+x_{j,1}-1 \quad  \mbox{for} \quad 1 \leq i < j \leq s
\end{array}
\right\} .
\end{displaymath}

Let
\begin{equation}
\EuScript{M}^\circ(\lambda, L) := \sum_{\boldsymbol{x} \in \Omega^\circ (\lambda, L)} 
q^{E_{\rm RC}(\lambda,\boldsymbol{x})},
\end{equation}
where $E_{\rm RC}(\lambda,\boldsymbol{x})$ is given in Definition \ref{def:may12_4}.
In section \ref{sec:5} we will show:
\begin{lemma}\label{lem:oct19_2}
The following relation holds:
\begin{equation}\label{eq:nov1_2}
\EuScript{M}^\circ(\lambda, L) =
q^{\psi (\lambda)} \frac{[L]}{[p_1]}
\prod_{i=1}^{s}
\left[
\begin{array}{c}
p_i+m_i-1\\
m_i
\end{array}
\right].
\end{equation}
In case $p_1 = 0$, the combination $\frac{[L]}{[p_1]} {p_1+m_1-1 \brack m_1} $
is to be understood as $\frac{[L]}{[m_1]}$.
\end{lemma}

Let $P^\circ_{L,\lambda} = \Phi^{-1}(\{ \lambda \} \times \Omega^\circ(\lambda,L))$,
where $\Phi$ is a bijective map to be defined in section \ref{subsec:4_2}.
Hence $|P^\circ_{L,\lambda}| = |\{ \lambda \} \times \Omega^\circ(\lambda,L)| = |\Omega^\circ(\lambda,L)|$.
Note that RHS of (\ref{eq:nov1_2}) reduces to that of (\ref{eq:nov1_1}) 
when $q = 1$,
implying $|P^\circ_{L,\lambda}| = |{\mathcal P}^\circ_{L,\lambda}|$.
In fact we have
\begin{proposition}\label{pr:new}
$P^\circ_{L,\lambda} = {\mathcal P}^\circ_{L,\lambda}$.
\end{proposition}
\textit{Proof.}
This is due to the second definition of $P^\circ_{L,\lambda}$
based on the KKR bijection explained previously, 
the footnote on the quantum numbers in section \ref{sec:2},
Remark \ref{re:mit} and 
Theorem 4.3 of \cite{KS}.
\hfill $\Box$

\begin{example}\label{ex:oct19_1}
Let $L=6,| \lambda |=3$. Then
\begin{eqnarray*}
P^\circ_{6, (3)} &=& \{ 222111, 122211, 112221, 111222, 211122, 221112 \}, \\
P^\circ_{6, (111)} &=& \{ 212121, 121212 \},\\
P^\circ_{6, (21)} &=& \{ 212211, 221211, 211221, 221121, \\
&& \, 121221, 211212, 212112, 122121, \\
&& \, 121122, 122112, 112122, 112212 \}.
\end{eqnarray*}
\end{example}
By Theorem \ref{th:oct20_1} we have
$\bigsqcup_{\lambda \vdash M} P^\circ_{L,\lambda} = P_{L,M}$.
We define
\begin{equation}
X^\circ (\lambda, L) := \sum_{p \in P^\circ_{L,\lambda}} 
q^{E_{\rm path}(p)}.
\end{equation}

\begin{example}
By Example \ref{ex:oct19_1} it is easy to check
\begin{eqnarray*}
X^\circ ((3),6) &=& 1 +q + q^2 + q^3 + q^4 + q^5, \\
X^\circ ((111),6) &=& q^6 + q^9, \\
X^\circ ((21),6) &=& q^2 +2 q^3 + 2 q^4 + 2 q^5 + 2 q^6 + 2 q^7 + q^8.
\end{eqnarray*}
\end{example}

As a result of Theorems \ref{th:oct20_1}, \ref{th:oct20_2} 
to be shown in subsection \ref{subsec:4_4},
and Lemma \ref{lem:oct19_2}, we obtain the main result of this paper:
\begin{theorem}\label{th:main}
\begin{equation*}
X^\circ (\lambda, L) = \EuScript{M}^\circ (\lambda, L).
\end{equation*}
\end{theorem}

\begin{example}
By Lemma \ref{lem:oct19_2} we obtain
\begin{eqnarray*}
\EuScript{M}^\circ ((3),6) &=& q^0 \frac{[6]}{[1]} = 1 +q + q^2 + q^3 + q^4 + q^5, \\
\EuScript{M}^\circ ((111),6) &=& q^6 \frac{[6]}{[3]} = q^6  + q^9, \\
\EuScript{M}^\circ ((21),6) &=& q^2 \frac{[6]}{[1]} {2+1-1 \brack 1} = q^2 +2 q^3 + 2 q^4 + 2 q^5 + 2 q^6 + 2 q^7 + q^8.
\end{eqnarray*}
\end{example}

\section{A statistics preserving bijection between 
periodic paths and rigged configurations}\label{sec:4}
\subsection{Multisets}\label{subsec:4_1}
We present an algorithm for a 
bijection between
periodic paths and rigged configurations.
The map from the former to the latter is referred to as
a {\em direct scattering transform},
and its inverse is called an {\em inverse scattering transform}.
To describe them, we first introduce the notion of multisets.

\vspace{3mm}
A multiset is a set with repeated elements which are assumed to be integers throughout this paper.
When treating multisets, we respect
the multiplicity of their elements.
For instance $\{1,2,2,3\} \cap \{2,2,4,5\}$ is
equal to $\{2,2\}$, not to $\{2\}$.
And $\{1,2,2,3\} \setminus \{2,3\}$ is
equal to $\{1,2\}$, not to $\{1\}$.
Also we always rearrange elements of a multiset
in weakly increasing order
and call them its first, second, ..., last elements.
For instance 
$\{1,2,2,3\} \cup \{2,2,4,5\} = \{ 1,2,2,2,2,3,4,5 \}$, and its sixth element
is $3$.

\vspace{3mm}
We define ${\mathcal O}_0, {\mathcal O}_1$ to be a pair of operators acting
on multisets:
Given any multiset $M$, the operator ${\mathcal O}_0$ (resp.~${\mathcal O}_1$) 
adds $2i-2$ (resp.~$2i-1$) to its $i$-th element
for all $i \in \{ 1, \ldots , |M| \}$. 
For instance, ${\mathcal O}_0 (\{ 1,2,2,3 \}) = \{1,4,6,9 \}$ and
${\mathcal O}_1 (\{ 1,2,2,3 \}) = \{2,5,7,10 \}$.
Their inverse operators ${\mathcal O}_a^{-1} \, (a=0,1)$ are defined by an obvious way.

\subsection{direct scattering transform}\label{subsec:4_2}
The direct scattering transform is a map that sends a path $p = b_1 \ldots b_L$ to the following data:
a list of positive integer pairs $\lambda = \{ (\ell_1, m_1), \ldots, (\ell_s, m_s) \}$ (soliton content)
and an integer vector
$\boldsymbol{x} =(\boldsymbol{x}_{i})_{1 \leq i \leq s}, \boldsymbol{x}_i =(x_{i,j})_{1 \le j \le m_{i}}$
obeying the condition $x_{i,1} \leq \cdots \leq x_{i,m_i}$ (angle variables).
Denote this map by $\Phi: p \mapsto (\lambda, \boldsymbol{x})$.

The path $p$ can be represented by a pair of multisets $M_1, M_2$
constructed by the following procedure:
Let $M_1=\{0 \leq j \leq L\mid (b_j, b_{j+1}) =(1, 2)\}$
and $M_2=\{0 \leq j \leq L\mid (b_j, b_{j+1}) =(2,1)\}$ 
(arranged in weakly increasing order),
where $b_0 = b_{L+1}=1$.
Since $|M_1|=|M_2|$ one can write them as $M_1 = \{ \alpha_1, \ldots, \alpha_N \}$ and $M_2 = \{ \beta_1, \ldots, \beta_N \}$
for some $N \in \Z_{>0}$.
At the beginning their elements satisfy
the conditions $0 \leq \alpha_1 < \beta_1 < \cdots < \alpha_N < \beta_N \leq L$ 
and
$(\beta_1 - \alpha_1) + \cdots +  (\beta_N - \alpha_N) \leq L/2$.

\vspace{3mm}
Let $M_1, M_2$ be the multisets for a path $p$ of length $L$.
Introduce a pair of indices $j,k$, a list $D$, and a multiset $S$.
To begin with we set $j=0, k = L$, $D = \{ \, \}$ (an empty list), and $S = \emptyset$.
Then run the following algorithm:

\begin{description}
\item[DS0] If $0 \in M_1$ and $ L \in M_2$, 
then replace $M_1$ by $M_1 \setminus \{0\}$ and
$M_2$ by $(M_2 \setminus \{\min M_2, L\}) \cup \{ \min M_2 +L \}$.
\item[DS1] While $M_1 \cap M_2 = \emptyset$ and {$\max M_2 - \min M_1 < k$}, 
continue replacing $(j,k,M_1,M_2)$ by $(j+1,k-2|M_1|,{\mathcal O}_0^{-1}(M_1),{\mathcal O}_1^{-1}(M_2))$.
\item[DS2] If $M_1 \cap M_2 \ne \emptyset$, then
set $S = M_1 \cap M_2$ and
replace $(M_1, M_2)$ by $(M_1 \setminus S, M_2 \setminus S)$.
\item[DS3] If $\max M_2 - \min M_1 = k$, then
replace $S$ by $S \cup \{ \min M_1 \}$,
$M_1$ by $M_1 \setminus \{ \min M_1 \}$, and 
$M_2$ by {$(M_2 \setminus \{\min M_2, \max M_2 \}) \cup \{\min M_2 + k \}$}.
\item[DS4] Pre-pend (i.e. append as the first element) $\{ S, j, k \}$ to $D$ and set $S = \emptyset$.
If $M_1 = \emptyset$ then stop.
Otherwise go to step \textbf{DS1}.
\end{description}
At the end we obtain such type of data
$D=\{ \{S_1,j_1, k_1 \}, \{S_2,j_2, k_2\},\ldots,\{S_s,j_s, k_s\} \}$
for some $s$, where $S_1, \ldots, S_s$ are multisets,
$j_1, \ldots, j_s, k_2, \ldots, k_s$ are positive integers, and $k_1$ is a non-negative integer.

\vspace{3mm}
Let $\ell_i = j_i, p_i = k_i$,
$m_i = |S_i|$ and
$\boldsymbol{x}_i = S_i$.

\vspace{3mm}
We show that this algorithm is well-defined and reversible, as well as that the condition $\ell_1 > \ldots > \ell_s > 0$
holds.
The reverse algorithm ( \textbf{IS1} -- \textbf{IS4}) will be given in the next subsection.
Write the $M_1, M_2$ as
$M_1 = \{ \alpha_1, \ldots, \alpha_N \}$ and $M_2 = \{ \beta_1, \ldots, \beta_N \}$
although their elements and even their cardinality $N$ will be repeatedly updated while the algorithm is running.
Note that the following inequity is always satisfied:
\begin{equation}\label{eq:aug25_1}
(\beta_1 - \alpha_1) + \cdots +  (\beta_N - \alpha_N) \leq k/2.
\end{equation}

\vspace{5mm}

\textit{About step \textbf{DS0}}:
When $\alpha_1 = 0$ and $\beta_N = L$ we redefine $M_1, M_2$ as
$M_1 = \{ \alpha_2, \ldots, \alpha_N \}, M_2 = \{ \beta_2, \ldots, \beta_{N-1}, \beta_1 + L \}$.
One has $\max M_2 > L$ when this action has been taken, while otherwise $\max M_2 \leq L$.
This implies that we can retrieve original $M_1, M_2$ by taking step \textbf{IS4} in the reverse algorithm.

\vspace{5mm}

\textit{About step \textbf{DS1}}:
Every replacement in step \textbf{DS1} reduces $\beta_i - \alpha_i, \alpha_{i+1} - \beta_i$ and $k-(\beta_N - \alpha_1)$ by one.
Thus
$M_1 \cap M_2 \ne \emptyset$ or $\max M_2 - \min M_1 = k$ will be certainly attained
after taking this action finite times.
Since no rearrangements of elements occur during this step,
one can go backwards and retrieve original $M_1, M_2$ from their final expressions in this step.

\vspace{5mm}

\textit{About steps \textbf{DS2} and \textbf{DS3}}:

A reverse procedure for these steps is given by \textbf{IS1} and \textbf{IS2} in the algorithm of the inverse scattering transform.
Suppose step \textbf{DS3} is relevant.
(The other case is easier.
If \textbf{DS3} is irrelevant, then so is \textbf{IS2} in the reverse algorithm.
It will be also shown that 
there is a condition on the expressions of
the $M_1$ and $M_2$ just after \textbf{DS3}
which enables us to distinguish whether step \textbf{DS3} has been relevant or not.)
Then the $M_1, M_2$ which one has just before step \textbf{DS2} can be written as
\begin{equation}\label{eq:aug4_3}
M_1 = \{ \alpha_1, \ldots, \alpha_N \}, M_2 = \{ \beta_1, \ldots, \beta_N = \alpha_1 + k \}.
\end{equation}
In order to leave an $\alpha_1$ in $M_1$ just after step \textbf{DS2} there must be odd number of $\alpha_1$'s 
in $M_1 \cup M_2$.
Thus one has
\begin{eqnarray*}
&&\alpha_1 = \beta_1  = \cdots = \alpha_{p-1} = \beta_{p-1} = \alpha_p < \beta_p < \alpha_{p+1}, \quad \mbox{or} \\
&&\alpha_1 = \beta_1  = \cdots = \alpha_{p-1}= \beta_{p-1} = \alpha_p < \beta_p = \alpha_{p+1} \leq
\cdots \leq \beta_{q-1} = \alpha_{q} \leq \beta_q < \alpha_{q+1}, 
\end{eqnarray*} 
for some $p,q \, (1 \leq p < q)$.
Consider the latter case.
(The former case is easier.)
Write the $M_1, M_2$ which one has just after step \textbf{DS2} as
$M_1 = \{ \alpha'_1, \ldots, \alpha'_M \}$ and $M_2 = \{ \beta'_1, \ldots, \beta'_M \}$ where
$\alpha'_1 = \alpha_1, \beta'_1=\beta_q$, and
$\beta'_M =  \alpha_1 + k$.
Assuming $M=1$ one has $\beta'_1 - \alpha'_1 = k > k/2 $, 
which contradicts (\ref{eq:aug25_1}).
Hence $M \geq 2$.
Now we get
\begin{equation}\label{eq:aug4_1}
M_1 = \{ \alpha'_2, \ldots, \alpha'_M \}, M_2 = \{ \beta'_2, \ldots, \beta'_{M-1}, \beta_q + k \}, 
\end{equation}
and
\begin{equation}\label{eq:aug4_2}
S = \{ \overbrace{\alpha_1, \ldots , \alpha_1}^{p}, \alpha_{p+1}, \ldots, \alpha_{q}, \ldots \},
\end{equation}
just after step \textbf{DS3}.
Note that $\alpha_2' \geq \alpha_{q+1}$.

Now let us consider how one can retrieve the original $M_1, M_2$.
Starting from (\ref{eq:aug4_1}) and (\ref{eq:aug4_2}), 
we replace $(M_1, M_2)$ by $(M_1 \cup S, M_2 \cup S)$ according to \textbf{IS1}.
Then 
\begin{eqnarray}
&& M_1 = \{ \overbrace{\alpha_1, \ldots , \alpha_1}^{p}, \alpha_{p+1}, \ldots, \alpha_{q}, 
\alpha_{q+1}, \ldots, \alpha_{N-1}, \alpha_{N} \},\label{eq:aug31_1}\\
&& M_2 = \{ \overbrace{\alpha_1, \ldots , \alpha_1}^{p}, \alpha_{p+1}, \ldots, \alpha_{q}, 
\beta_{q+1}, \ldots, \beta_{N-1}, \beta_q + k \}.\label{eq:aug31_2}
\end{eqnarray}
Hence the $M_1$ in (\ref{eq:aug4_3}) has been retrieved.
Then applying \textbf{IS2} we get the $M_2$ in (\ref{eq:aug4_3}) by replacing
$M_2$ by $(M_2 \setminus \{ \min M_2, \max M_2 \}) \cup \{ \max M_2 -k,  \min M_2 + k\}$.
Note that $\max M_2 - \min M_1 > k$ is satisfied for (\ref{eq:aug31_1}),(\ref{eq:aug31_2}), while if \textbf{DS3} is 
irrelevant this does not hold.
This is the condition that we have mentioned just above (\ref{eq:aug4_3}).
%

\vspace{5mm}

\textit{About the condition $\ell_1 > \ldots > \ell_s > 0$}:

We note that for (\ref{eq:aug4_1}) one has
$\max M_2 - \min M_1 = \beta_q + k - \alpha'_2 \leq \beta_q + k - \alpha_{q+1} < k$.
Hence whether step \textbf{DS3} is relevant or not, we have
$M_1 \cap M_2 = \emptyset$ and $\max M_2 - \min M_1  < k$ just after step \textbf{DS3}.
These conditions are also valid just after step \textbf{DS0}.
Thus the increment $j \rightarrow j+1$ in
the procedure in step \textbf{DS1} is certainly done at least once whenever the algorithm takes this step,
leading to the condition $\ell_1 > \ldots > \ell_s > 0$.

\begin{example}
Consider the path $p=21221221111221$ of length $L=14$.
Then $M_1 = \{ 0,2,5,11 \}$ and $M_2 = \{ 1,4,7,13 \}$.
By applying the procedure in step \textbf{DS1} once,
we obtain
$j=1, k=14-8=6, M_1 = \{ 0,0,1,5 \}$ and $M_2 = \{ 0,1,2,6 \}$.
By \textbf{DS2} we set $S = \{ 0, 1 \}$ and redefine $M_1$ as $M_1 = \{ 0,5 \}$, 
$M_2$ as $M_2 = \{ 2,6 \}$.
By \textbf{DS3} we set $S = \{ 0, 0, 1 \}$ and redefine $M_1$ as $M_1 = \{ 5 \}$, 
$M_2$ as $M_2 = \{ 8 \}$.
Then by \textbf{DS4}
we have $D=\{ \{ \{ 0,0,1 \},1, 6\} \}$, and go to \textbf{DS1}.

In the second turn, 
we obtain 
$j=4, k=6-3 \times 2 =0, M_1 = M_2 = \{ 5 \}$ after applying the procedure in \textbf{DS1} three times.
By \textbf{DS2} we set $S = \{ 5 \}$ and redefine $M_1$ and $M_2$ as $M_1 = M_2 =  \emptyset$. 
Then by \textbf{DS4}
we have $D=\{\{ \{ 5 \},4,0 \} , \{ \{ 0,0,1 \},1,6\}  \}$, and stop the algorithm.

Hence we have $(\ell_1,\ell_2) = ( 4,1 ), (p_1,p_2)= (0,6), (m_1,m_2)= (1,3)$ and
\begin{math}
\boldsymbol{x} =(x_{1,1},x_{2,1},x_{2,2},x_{2,3}) = (5,0,0,1).
\end{math}
\end{example}

\begin{example}
Consider the path $p=22211222111111221$ of length $L=17$.
Then $M_1 = \{ 0,5,14 \}$ and $M_2 = \{ 3,8,16 \}$.
By applying the procedure in \textbf{DS1} once,
we obtain
$j=1, k=17-6=11, M_1 = \{ 0,3,10 \}$ and $M_2 = \{ 2,5,11 \}$.
By \textbf{DS3} we set $S = \{ 0 \}$ and redefine $M_1$ as $M_1 = \{ 3,10 \}$, 
$M_2$ as $M_2 = \{ 5,13 \}$.
Then by \textbf{DS4}
we have $D=\{ \{ \{ 0 \},1, 11\} \}$, and go to \textbf{DS1}.

In the second turn, 
we obtain 
$j=2, k=11-4 =7, M_1 = \{ 3,8 \}$ and $M_2 = \{ 4,10 \}$ after applying the procedure in \textbf{DS1} once.
By \textbf{DS3} we set $S = \{ 3 \}$ and redefine $M_1$ as $M_1 = \{ 8 \}$, 
$M_2$ as $M_2 = \{ 11 \}$.
Then by \textbf{DS4}
we have $D=\{\{ \{ 3 \},2,7\} , \{ \{ 0 \},1,11\}  \}$, and go to \textbf{DS1}.

In the third turn, 
we obtain 
$j=5, k=7-3 \times 2 =1, M_1 = M_2  = \{ 8 \}$ after applying the procedure in \textbf{DS1} three times.
Then by \textbf{DS2} and \textbf{DS4}
we have $D=\{\{ \{ 8 \},5,1\} ,  \{ \{ 3 \},2,7\} , \{ \{ 0 \},1,11\}  \}$, 
$M_1 = M_2 = \emptyset$, and stop the algorithm.

Hence we have $(\ell_1,\ell_2,\ell_3) = ( 5,2,1 ), (p_1,p_2,p_3)= (1,7,11), (m_1,m_2,m_3)= (1,1,1)$ and
\begin{math}
\boldsymbol{x} =(x_{1,1},x_{2,1},x_{3,1}) = (8,3,0).
\end{math}
\end{example}

\begin{example}
Consider the path $p=22221111222211111222111211122211122221$ of length $L=38$.
Then $M_1 = \{ 0,8,17,23,27,33 \}$ and $M_2 = \{ 4,12,20,24,30,37 \}$.
By applying the procedure in \textbf{DS1} once,
we obtain
$j=1, k=38-12=26, M_1 = \{ 0,6,13,17,19,23 \}$ and $M_2 = \{ 3,9,15,17,21,26 \}$.
By \textbf{DS2} we set $S = \{ 17 \}$ and redefine $M_1$ as $M_1 = \{ 0,6,13,19,23 \}$,
$M_2$ as $M_2 = \{ 3,9,15,21,26 \}$.
By \textbf{DS3} we redefine $S$ as $S = \{ 0, 17\}$, 
$M_1$ as $M_1 = \{ 6,13,19,23 \}$,
$M_2$ as $M_2 = \{ 9,15,21,29 \}$.
Then by \textbf{DS4}
we have $D=\{ \{ \{ 0, 17 \},1, 26\} \}$, and go to \textbf{DS1}.

In the second turn, 
we obtain 
$j=3, k=26-2 \times 8 =10, M_1 = \{ 6,9,11,11 \}$ and $M_2 = \{ 7,9,11,15 \}$ after applying the procedure in \textbf{DS1} twice.
Then by \textbf{DS2} and \textbf{DS4} one has $D=\{\{ \{ 9, 11 \},3,10\} , \{ \{ 0, 17 \},1,26\}  \}$,
$M_1 = \{ 6,11 \}, M_2 = \{ 7,15 \}$, and go to \textbf{DS1}.

After two more turns, 
one has
\begin{equation*}
D =\{\{ \{ 9 \},7,0\} , \{ \{ 6 \},4,6\} , \{ \{ 9, 11 \},3,10\} , \{ \{ 0, 17 \},1,26\}  \},
\end{equation*} 
$M_1 =  M_2 = \emptyset$, and stop the algorithm.
Hence we have $(\ell_1,\ell_2,\ell_3,\ell_4) = ( 7,4,3,1 ), (p_1,p_2,p_3,p_4)= (0,6,10,26), (m_1,m_2,m_3,m_4)= (1,1,2,2)$ and
\begin{equation*}
\boldsymbol{x} =(x_{1,1},x_{2,1},x_{3,1},x_{3,2},x_{4,1},x_{4,2}) = (9,6,9,11,0,17).
\end{equation*}
\end{example}

\begin{remark}\label{rem:july28_1}
By construction the relations $p_1 = L - 2 \sum_{i=1}^s \ell_i m_i$ and
$p_i = p_{i-1} + 2 (m_1 + \cdots + m_{i-1}) (\ell_{i-1} - \ell_i) \quad (2 \leq i \leq s)$ hold.
Thus the calculation of the vacancy numbers in the algorithm is optional.
%
We assumed that the number of $1$'s is not less than that of $2$'s
for any paths, which implies $p_1 \geq 0$.
In addition, the condition $\ell_1 > \ldots > \ell_s >0$ holds as we mentioned.
Thus we have $0 \leq p_1 < \cdots < p_s (< L)$.
\end{remark}

\begin{remark}\label{re:mit}
Our $\Phi$ is equivalent to the map introduced 
in \cite{MIT} under the name of `10-elimination'.
The elements of $\boldsymbol{x}$ in this paper coincide with the data 
called `the positions of 0-solitons'  in \cite{MIT}. 
In terms of the 10-elimination, 
the vacancy numbers represent effective sizes of the system.
\end{remark}


\subsection{Inverse scattering transform}\label{subsec:4_3}
The inverse scattering transform is defined as the inverse of $\Phi$
that sends soliton content $\lambda = \{ (\ell_1, m_1), \ldots, (\ell_s, m_s) \}$ 
and angle variable 
$\boldsymbol{x} = (\boldsymbol{x}_i)_{1 \leq i \leq s}$ to a path $p$.
Denote this map by $\Phi^{-1}: (\lambda, \boldsymbol{x}) \mapsto p$.

Let $S_i = \boldsymbol{x}_i$ interpreted as a multiset, and $p_i = p_i (\lambda)$ be the vacancy number.
Essentially, the inverse scattering transform is a map that sends data 
$D=\{ \{S_1,\ell_1, p_1 \}, \{S_2,\ell_2, p_2\},\ldots,\{S_s,\ell_s, p_s\} \}$
to a pair of multisets $M_1, M_2$.
Now we describe it based on the discussion on the reversibility of $\Phi$
in the previous subsection. 

To begin with we set $M_1 = M_2 = \emptyset$ and $i=1$.
Then run the following algorithm.
\begin{description}
\item[IS1] Replace $M_1$ by $M_1 \cup S_i$ and $M_2$ by $M_2 \cup S_i$.
\item[IS2] If $\max M_2 - \min M_1 > p_i$, then replace $M_2$ by 
$(M_2 \setminus \{ \min M_2, \max M_2 \}) \cup \{ \max M_2 -p_i,  \min M_2 + p_i\}$.
\item[IS3] Replace $M_1$ by ${\mathcal O}_0^{\ell_i-\ell_{i+1}} (M_1)$ and $M_2$ by ${\mathcal O}_1^{\ell_i-\ell_{i+1}} (M_2)$.
If $i < s$, then replace $i$ by $i+1$ and go back to step \textbf{IS1}.
If $i = s$, then go to step \textbf{IS4} where we interpret $\ell_{s+1} = 0$.
\item[IS4] If $\max M_2 > L$, then replace $M_1$ by $M_1 \cup \{ 0 \}$
and $M_2$ by 
$(M_2 \setminus \{ \max M_2 \}) \cup \{ \max M_2 -L, L\}$.
\end{description}
%

Path $p=b_1 \cdots b_L  \in P_{L,M}$ can be represented by multisets
$M_1 = \{ \alpha_1, \ldots, \alpha_N \}$ and $M_2 = \{ \beta_1, \ldots, \beta_N \}$ for some $N$.
Their elements obey the following conditions: 
\begin{eqnarray}
&& 0 \leq \alpha_1 < \beta_1 < \cdots < \alpha_N < \beta_N \leq L, \label{eq:aug18_3} \\
&& \sum_{i=1}^N (\beta_i - \alpha_i) = M. \label{eq:aug18_4}
\end{eqnarray}
To obtain $p \in P_{L,M}$ from $M_1, M_2$, set $b_i = 1$ (resp.~$b_i=2$) if $\beta_{k-1} < i \leq \alpha_k$
(resp.~$\alpha_{k} < i \leq \beta_k$) for some $k$.
Here we interpret $\beta_{-1}=0$ and $\alpha_{N+1} = L$.
Since $\Phi$ is reversible, it gives a bijection between $P_{L,M}$ and $\Phi(P_{L,M})$.
In the next subsection 
we classify the elements of $\Phi (P_{L,M})$ according to their soliton contents.
For this purpose we establish a lemma in this subsection.

Let $\lambda = \{ (\ell_i, m_i) \}_{1 \leq i \leq s}$ be a soliton content and
recall the set 
$\Omega^\circ(\lambda,L)$
defined in subsection \ref{subsec:3_2}
where the angle variables
$\boldsymbol{x} =(\boldsymbol{x}_{i})_{1 \leq i \leq s}, \boldsymbol{x}_i =(x_{i,j})_{1 \le j \le m_{i}}$
are satisfying the conditions:
\begin{eqnarray}
&& 0 \leq x_{i,1} \leq \cdots \leq x_{i,m_i} \quad \mbox{for} \quad 1 \leq i \leq s, \label{eq:july29_2}\\
&& x_{i,m_i} \leq p_i + \min (x_{i,1},2 \ell_i-1) \quad \mbox{for} \quad 1 \leq i \leq s, \label{eq:july29_3}\\
&& x_{j,m_j} - p_j \leq x_{i,1} \quad \mbox{for} \quad 1 \leq i < j \leq s, \label{eq:july29_4}\\
&& x_{i,m_i} \leq p_i+2 (\ell_i-\ell_j)+x_{j,1}-1 \quad \mbox{for} \quad 1 \leq i < j \leq s.\label{eq:july29_5}
\end{eqnarray}

\begin{lemma}\label{lem:aug6_2}
For any $\lambda \vdash M$, the condition $\{ \lambda \} \times \Omega^\circ(\lambda,L) \subset \Phi (P_{L,M})$ holds.
\end{lemma}
\noindent
\textit{Proof.}
Given $\lambda \vdash M$ and $\boldsymbol{x} \in \Omega^\circ(\lambda,L)$, let $D$ be the data associated with $(\lambda, \boldsymbol{x})$.
We are to show that the algorithm of $\Phi^{-1}$ is well-defined for such $D$, and
the resulting $M_1, M_2$ satisfy the conditions (\ref{eq:aug18_3}), (\ref{eq:aug18_4}).

Recall the multisets $M_1, M_2$ used in the algorithm.
Write them as
$M_1 = \{ \alpha_1, \ldots, \alpha_N \}$ and $M_2 = \{ \beta_1, \ldots, \beta_N \}$
although their elements and even
their cardinality $N$ will be repeatedly updated while the algorithm is running.
It will be shown that the following conditions are satisfied:
\begin{enumerate}
\item $\alpha_1 \leq \beta_1 \leq \cdots \leq \alpha_N \leq \beta_N$ and $\alpha_N - \alpha_1 \leq p_i$,
just after step \textbf{IS1}.
\item $\alpha_1 \leq \beta_1 \leq \cdots \leq \alpha_N \leq \beta_N$ and $\beta_N - \alpha_1 \leq p_i$,
just after step \textbf{IS2}.
\item $\alpha_1 < \beta_1 < \cdots < \alpha_N < \beta_N$ and $\beta_N - \alpha_1 < p_{i+1}$,
just after step \textbf{IS3}.
\item $0 \leq \alpha_1$ and $\alpha_N \leq L-1$, just before step \textbf{IS4}.
\end{enumerate}
Here we interpret
$p_{s+1} = p_{s} + 2 (m_1 + \cdots + m_{s}) \ell_{s} = L$.
We shall prove (i)-(iv) later.
The statements of items (i)-(iii) are for the $i$-th turn of the algorithm.
We denote by (x)$_k$ the statement of item (x) for the $k$-th turn.
Items (i)$_1$, (ii)$_1$, and (iv) can be proved independently.
We shall show that: 1. item (i)$_{k > 1}$ holds under the assumption of item (iii)$_{k-1}$;
2. item (ii)$_{k > 1}$ holds under (iii)$_{k-1}$ and (i)$_{k}$; 
3. item (iii)$_{k}$ holds under (ii)$_{k}$.
Thus (i)$_{k}$-(iii)$_{k}$ can be proved by induction on $k$.

The Lemma follows from items (iii) and (iv).
Denote those $M_1$ and $M_2$ that one has just before step \textbf{IS4} by
$M'_1 = \{ \alpha'_1, \ldots, \alpha'_N \}$ and $M'_2 = \{ \beta'_1, \ldots, \beta'_N \}$.
Then we have $0 \leq \alpha'_1 < \beta'_1 < \ldots < \alpha'_N < \beta'_N$,
$\beta'_N - \alpha'_1 < p_{s+1}= L$ and $\alpha'_N \leq L-1$ by items (iii) and (iv).
We can also derive
\begin{equation*}
\sum_{i=1}^N (\beta'_i - \alpha'_i) =
\sum_{i=1}^s (m_1 + \cdots + m_i) (\ell_i - \ell_{i+1})=
\sum_{i=1}^s m_i \ell_i = |\lambda|,
\end{equation*}
based on step \textbf{IS3} where the relation 
$|M_1| = |M_2| = m_1 + \cdots + m_i$ holds in the $i$-th turn.

We are to show that the $M_1, M_2$ that one has after step \textbf{IS4} satisfy the conditions
(\ref{eq:aug18_3}), (\ref{eq:aug18_4}) with $M = |\lambda|$.
If $\beta'_N \leq L$ then step \textbf{IS4} is irrelevant and 
the Lemma follows immediately.

Suppose $\beta'_N > L$ and 
write the $M_1, M_2$ that one has after step \textbf{IS4} as
$M_1 = \{ \alpha_1, \ldots, \alpha_{N+1} \}, M_2 = \{ \beta_1, \ldots, \beta_{N+1} \}$.
Since $M_1 = M'_1 \cup \{ 0 \}$ and $\alpha'_1 \geq 0$, 
we have $\alpha_1 = 0$ and $\alpha_i = \alpha'_{i-1} \, (2 \leq i \leq N+1)$ for the elements of $M_1$.
While the $M_2$ is given in terms of the elements of $M_2'$ by
$M_2 = \{ \beta'_1, \ldots, \beta'_{N-1} \} \cup \{ \beta'_N -L, L \}$.
By assumption we have $\beta'_N -L < \alpha'_1 < \beta'_1$, and $L > \alpha'_N > \beta'_{N-1}$.
Hence $\beta_1 = \beta'_N -L, \beta_{N+1} = L$, and $\beta_i = \beta'_{i-1} \, (2 \leq i \leq N)$.
Now it is easy to see that the elements of $M_1, M_2$ obey 
the conditions 
$0 = \alpha_1 < \beta_1 < \ldots < \alpha_{N+1} < \beta_{N+1} = L$
and 
$\sum_{i=1}^{N+1} (\beta_i - \alpha_i) = \sum_{i=1}^{N} (\beta'_i - \alpha'_i) = |\lambda|$.
\vspace{5mm}

\textit{Proof of item (i)}:
When $i=1$ we have $M_1 = M_2 = S_1 = \{ x_{1,1}, \ldots , x_{1,m_1} \}$.
Hence the claim follows by (\ref{eq:july29_2}) and (\ref{eq:july29_3}).
Suppose $i > 1$.
Denote by
$M'_1 = \{ \alpha'_1, \ldots, \alpha'_M \}$ and $M'_2 = \{ \beta'_1, \ldots, \beta'_M \}$
the $M_1, M_2$ that one has just after step \textbf{IS3} in the $(i-1)$-th turn.
We assume the conditions
$\alpha'_1 < \beta'_1 < \ldots < \alpha'_M < \beta'_M$ and $\beta'_M - \alpha'_1 < p_{i}$. 
Let $M_1 = \{ \alpha_1, \ldots, \alpha_N \}$ and $M_2 = \{ \beta_1, \ldots, \beta_N \}$
be the $M_1, M_2$ that one has just after step \textbf{IS1} in the $i$-th turn.
Then $M_1 = M'_1 \cup S_i$ and $M_2 = M'_2 \cup S_i$ with $S_i = \{ x_{i,1}, \ldots , x_{i,m_i} \}$.
Clearly the condition
$\alpha_1 \leq \beta_1 \leq \cdots \leq \alpha_N \leq \beta_N$ 
holds.
Let us show $\alpha_N - \alpha_1 \leq p_i$.
It suffices to consider the following four cases:
\begin{description}
\item[case 1] $\alpha_1 = x_{i,1}, \alpha_N = x_{i,m_i}$: 
By (\ref{eq:july29_3}) we have $\alpha_N - \alpha_1 = x_{i,m_i} - x_{i,1} \leq p_i$.
\item[case 2] $\alpha_1 = \alpha'_1, \alpha_N = \alpha'_M$:
Since $\beta'_M - \alpha'_1 < p_{i}$ and $\alpha'_M < \beta'_M$, we have
$\alpha_N - \alpha_1 = \alpha'_M - \alpha'_1 < \beta'_M - \alpha'_1 < p_{i}$.
\item[case 3] $\alpha_1 = \alpha'_1, \alpha_N = x_{i,m_i}$:
The $\alpha'_1$ is given as $\alpha'_1 = x_{j,1}$ for some $j (< i)$.
Hence 
$\alpha_N - \alpha_1 = x_{i,m_i} - x_{j,1} \leq p_{i}$ by (\ref{eq:july29_4}).
\item[case 4] $\alpha_1 = x_{i,1}, \alpha_N = \alpha'_M$:
The $\alpha'_M$ is given by the following formula for some $j (< i)$:
\begin{eqnarray*}
\alpha'_M &=& x_{j,m_j} + 2 \sum_{k=j}^{i-1} (\ell_k - \ell_{k+1}) (\sum_{h=1}^k m_h - 1) \\
&=& x_{j,m_j} + \sum_{k=j}^{i-1} \left( (p_{k+1} - p_{k}) - 2 (\ell_k - \ell_{k+1}) \right) \\
&=& x_{j,m_j} + p_i - p_j + 2 \ell_i - 2 \ell_j.
\end{eqnarray*}
Hence 
$\alpha_N - \alpha_1 = x_{j,m_j} - x_{i,1} + p_i - p_j + 2 \ell_i - 2 \ell_j < p_{i}$ by (\ref{eq:july29_5}).
\end{description}
\vspace{5mm}

\textit{Proof of item (ii)}:
When $i=1$ step \textbf{IS2} is irrelevant, hence $M_1 = M_2 = S_1 = \{ x_{1,1}, \ldots , x_{1,m_1} \}$.
Thus the claim follows by (\ref{eq:july29_2}) and (\ref{eq:july29_3}).
Suppose $i > 1$.
Set $\alpha = \min M''_1$ and $\beta = \max M''_2$
where $M''_1, M''_2$ are the $M_1, M_2$ that one has just after step \textbf{IS3} in the $(i-1)$-th turn.
Assume $\beta - \alpha < p_i$.
We denote by
$M'_1 = \{ \alpha'_1, \ldots, \alpha'_N \}$ and $M'_2 = \{ \beta'_1, \ldots, \beta'_N \}$
the $M_1, M_2$ that one has just after step \textbf{IS1} in the $i$-th turn, and assume that the conditions
$\alpha'_1 \leq \beta'_1 \leq \ldots \leq \alpha'_N \leq \beta'_N$ and $\alpha'_N - \alpha'_1 \leq p_{i}$ are satisfied. 
Let $M_1 = \{ \alpha_1, \ldots, \alpha_N \}$ and $M_2 = \{ \beta_1, \ldots, \beta_N \}$
be the $M_1, M_2$ that one has just after step \textbf{IS2} in the $i$-th turn.
Since \textbf{IS2} keeps $M_1$ unchanged, one always has $\alpha_k = \alpha'_k \, (1 \leq k \leq N)$.
If $\beta'_N - \alpha'_1 \leq p_i$ then $\beta_k = \beta'_k \, (1 \leq k \leq N)$,
hence follows item (ii).

Suppose $\beta'_N - \alpha'_1 > p_i$.
In this case $\beta'_N = \beta$, because otherwise one has
 $\beta'_N = x_{i,m_i} = \alpha'_N$ and hence $\beta'_N - \alpha'_1 \leq p_i$.
Let $p$ be the smallest integer such that $\alpha'_p = \alpha$.
Then we have $p > 1$, because otherwise one has $\beta'_N - \alpha'_1 =\beta - \alpha < p_i$.
By definition of \textbf{IS1}, one has $\alpha'_k = \beta'_k$ for all $k \in \{ 1 \ldots, p-1 \}$.
By definition of \textbf{IS2},
one has $M_2 = \{ \beta'_2, \ldots, \beta'_{N-1} \} \cup \{ \beta'_1 + p_i, \beta'_N - p_i \}$.
Moreover we can show that
\begin{eqnarray}
\{ \beta_1, \ldots, \beta_{p-1} \} &=& \{ \beta'_2, \ldots, \beta'_{p-1} \} \cup \{ \beta'_N - p_i \}, \label{eq:aug18_1} \\
\{ \beta_p, \ldots, \beta_N \} &=& \{ \beta'_p, \ldots, \beta'_{N-1} \} \cup \{ \beta'_1 + p_i \}.\label{eq:aug18_2}
\end{eqnarray}
Since $\beta - \alpha < p_i$ one has
$\beta'_N - p_i = \beta - p_i < \alpha = \alpha'_p \leq \beta'_p$.
This implies (\ref{eq:aug18_1}).
Since $\alpha'_N - \alpha'_1 \leq p_{i}$ one has
$\beta'_1 + p_i \geq \beta'_1 +  \alpha'_N - \alpha'_1 = \alpha'_N \geq \beta'_{N-1}$.
This implies (\ref{eq:aug18_2}), or more precisely
$\beta_k = \beta'_k \, (p \leq k \leq N-1) $ and $\beta_N = \beta'_1 + p_i$.
Clearly the condition
$\alpha_p \leq \beta_p \leq \ldots \leq \alpha_N \leq \beta_N$ holds.
Noting that $\alpha'_k = \beta'_k$ for $1 \leq k \leq p-1$, it is also easy to see
$\alpha_1 \leq \beta_1 \leq \ldots \leq \alpha_{p-1} \leq \beta_{p-1}$.
Finally we find that
$\beta_{p-1} = \max (\beta'_{p-1}, \beta'_N - p_i) \leq \alpha'_p = \alpha_p$ and
$\beta_N - \alpha_1 = \beta'_1 + p_i - \alpha'_1 = p_i$,
completing the proof of item (ii).
\vspace{5mm}

\textit{Proof of item (iii)}:
Denote by
$M'_1 = \{ \alpha'_1, \ldots, \alpha'_N \}$ and $M'_2 = \{ \beta'_1, \ldots, \beta'_N \}$
the $M_1, M_2$ that one has just after step \textbf{IS2} in the $i$-th turn. 
We assume that the conditions
$\alpha'_1 \leq \beta'_1 \leq \ldots \leq \alpha'_N \leq \beta'_N$ and $\beta'_N - \alpha'_1 \leq p_{i}$ are satisfied. 
Let $M_1 = \{ \alpha_1, \ldots, \alpha_N \}$ and $M_2 = \{ \beta_1, \ldots, \beta_N \}$
be the $M_1, M_2$ that one has just after step \textbf{IS3} in the $i$-th turn.
Clearly the condition $\alpha_1 < \beta_1 < \cdots < \alpha_N < \beta_N$ holds.
It is also easy to see that
$\alpha_1 = \alpha'_1$
and
$\beta_N = \beta'_N + (\ell_i - \ell_{i+1})(2 \sum_{k=1}^i m_k -1) = \beta'_N + p_{i+1} - p_i + \ell_{i+1} - \ell_i$.
Hence
$\beta_N - \alpha_1 \leq p_{i+1} + \ell_{i+1} - \ell_i < p_{i+1}$.
\vspace{5mm}

\textit{Proof of item (iv)}:
Let $M_1 = \{ \alpha_1, \ldots, \alpha_N \}$
be the $M_1$ that one has just before step \textbf{IS4}.
The $\alpha_1$ is given as $\alpha_1 = x_{i,1}$ for some $i$.
Thus we have $\alpha_1 \geq 0$ by (\ref{eq:july29_2}).
The $\alpha_N$ is given by the following formula for some $i$:
\begin{eqnarray*}
\alpha_N &=& x_{i,m_i} + 2 \sum_{k=i}^{s} (\ell_k - \ell_{k+1}) (\sum_{h=1}^k m_h - 1) \\
&=& x_{i,m_i} + \sum_{k=i}^{s} \left( (p_{k+1} - p_{k}) - 2 (\ell_k - \ell_{k+1}) \right) \\
&=& x_{i,m_i} + L - p_i  - 2 \ell_i.
\end{eqnarray*}
Thus we have $\alpha_N \leq L-1$ by (\ref{eq:july29_3}).

\hfill $\Box$
\subsection{Statistics preserving bijection and the partition function}\label{subsec:4_4}
We classify the elements of $\Phi (P_{L,M})$ according to their
soliton contents.
As a result of Lemma \ref{lem:aug6_1} to be presented in the next section we have
\begin{equation*}
|\Omega^\circ(\lambda,L)| =
\frac{L}{m_1}
\left(
\begin{array}{c}
p_1+m_1-1\\
m_1-1
\end{array}
\right)
\prod_{i=2}^{s}
\left(
\begin{array}{c}
p_i+m_i-1\\
m_i
\end{array}
\right).
\end{equation*}
By using this expression one can show that $\sum_{\lambda \vdash M} |\Omega^\circ(\lambda,L)| = { L \choose M}$, while
Lemma \ref{lem:aug6_2} implies that $\bigsqcup_{\lambda \vdash M} \{ \lambda \} \times \Omega^\circ(\lambda,L) \subset \Phi (P_{L,M})$, 
which leads to
\begin{equation*}
\sum_{\lambda \vdash M} |\Omega^\circ(\lambda,L)| = |\bigsqcup_{\lambda \vdash M} \{ \lambda \} \times \Omega^\circ(\lambda,L) |
\leq |\Phi(P_{L,M})| =  |P_{L,M}|
=\left(
\begin{array}{c}
L \\
M
\end{array}
\right).
\end{equation*}
Hence $\bigsqcup_{\lambda \vdash M} \{ \lambda \} \times \Omega^\circ(\lambda,L) = \Phi (P_{L,M})$.
In other words we have: 
\begin{theorem}\label{th:oct20_1}
The map $\Phi$ gives a bijection between $P_{L,M}$ and $\bigsqcup_{\lambda \vdash M} \{ \lambda \} \times \Omega^\circ(\lambda,L)$. 
\end{theorem}
This bijection is {\em statistics preserving} in the following sense.

%

Given a path $p \in P_{L,M}$ let $M_1 = \{ \alpha_1 , \ldots, \alpha_N \}$ be its associated multiset, i.e.
for $p = b_1 \cdots b_L$ we
let $\alpha \in \{0 , \ldots,  L-1 \}$ belong to $M_1$ if and only if the condition $(b_\alpha, b_{\alpha+1}) =(1, 2)$
is satisfied, where we interpret $b_0 =1$.
Noting that $E_{\rm path}(p)= \sum_{i=1}^N \alpha_i$ it is easy to see that
the algorithm of the inverse scattering transform implies the following result.
\begin{theorem}\label{th:oct20_2}
Given $p \in P_{L,M}$ let $\Phi (p) = (\lambda, \boldsymbol{x})$.
Then $E_{\rm path}(p) = E_{\rm RC} (\lambda, \boldsymbol{x})$.
\end{theorem}
\textit{Proof}:
The energy $E_{\rm path}(p)$ is equal to sum of the elements of the multiset $M_1$. 
Consider how $M_1$ has been constructed in the algorithm of the inverse scattering transform.
One finds that step \textbf{IS1} is responsible for the term $\sum_{i=1}^s \sum_{j=1}^{m_i} x_{i,j}$ in
the definition of $E_{\rm RC} (\lambda, \boldsymbol{x})$, and
step \textbf{IS3} is for the $\psi (\lambda)$.
(See the way of its calculation presented above Example \ref{ex:nov2_1}.)
The other steps are irrelevant.
\hfill $\Box$

Let $P^\circ_{L,\lambda} = \Phi^{-1} (\{ \lambda \} 
\times \Omega^\circ(\lambda,L))$, which 
is the set of all paths of length $L$ and with soliton content $\lambda$.
Then $\Phi$ gives a statistics preserving bijection between $P^\circ_{L,\lambda}$ and $\{ \lambda \} \times \Omega^\circ(\lambda,L)$.
We consider the partition function for the periodic soliton cellular automaton over $P^\circ_{L,\lambda}$.
By Theorem \ref{th:oct20_2} we have
\begin{equation}
\sum_{p \in P^\circ_{L,\lambda}} q^{E_{\rm path}(p)} =
\sum_{\boldsymbol{x} \in \Omega^\circ(\lambda,L)} q^{E_{\rm RC}(\lambda, \boldsymbol{x})} =
q^{\psi (\lambda)} \sum_{\boldsymbol{x} \in \Omega^\circ(\lambda,L)} q^{|\boldsymbol{x}|},
\end{equation}
where $|\boldsymbol{x}| = \sum_{i=1}^s \sum_{k=1}^{m_i} x_{i,k}$.
In the next section we show that the sum $\sum_{\boldsymbol{x} \in \Omega^\circ(\lambda,L)} q^{|\boldsymbol{x}|}$
admits a fermionic expression.
\subsection{The case of non-periodic paths}\label{subsec:4_5}
In this subsection we briefly review direct and inverse scattering transforms for non-periodic paths \cite{Takagi1}.
We denote by $\Psi: p \mapsto (\lambda, \boldsymbol{x})$ the direct scattering transform, where $p$ is a path of length $L$.
Let $M_1, M_2, j, k, D, S$ be those introduced just above the algorithm \textbf{DS0}-\textbf{DS4} for the map $\Phi$
in subsection \ref{subsec:4_2}.
The algorithm for the map $\Psi$ is given as follows.
\begin{description}
\item[ds1] While $M_1 \cap M_2 = \emptyset$, 
continue replacing $(j,k,M_1,M_2)$ by $(j+1,k-2|M_1|,{\mathcal O}_0^{-1}(M_1),{\mathcal O}_1^{-1}(M_2))$.
\item[ds2] If $M_1 \cap M_2 \ne \emptyset$, then
set $S = M_1 \cap M_2$ and
replace $(M_1, M_2)$ by $(M_1 \setminus S, M_2 \setminus S)$.
\item[ds3] Pre-pend $\{ S, j \}$ to $D$ and set $S = \emptyset$.
If $M_1 = \emptyset$ then stop.
Otherwise go to step \textbf{ds1}.
\end{description}
At the end we obtain such type of data
$D=\{ \{S_1,j_1 \}, \{S_2,j_2 \},\ldots,\{S_s,j_s \} \}$
for some $s$, where $S_1, \ldots, S_s$ are multisets,
$j_1, \ldots, j_s $ are positive integers.
Let $\ell_i = j_i$, $m_i = |S_i|$ and
$\boldsymbol{x}_i = S_i$.
By Figure \ref{fig:1} the data $\{ (\ell_i , m_i) \}_{1 \leq i \leq s}$ determines a Young diagram $\lambda$.
\vspace{3mm}

Let $M_1, M_2, S_i$ be those introduced just above the algorithm \textbf{IS1}-\textbf{IS4} for the map $\Phi^{-1}$
in subsection \ref{subsec:4_3}.
The inverse map $\Psi^{-1}$ is given as follows.
To begin with we set $M_1 = M_2 = \emptyset$ and $i=1$.
Then run the following algorithm where we interpret $\ell_{s+1} = 0$.
\begin{description}
\item[is1] Replace $M_1$ by $M_1 \cup S_i$ and $M_2$ by $M_2 \cup S_i$.
\item[is2] Replace $M_1$ by ${\mathcal O}_0^{\ell_i-\ell_{i+1}} (M_1)$ and $M_2$ by ${\mathcal O}_1^{\ell_i-\ell_{i+1}} (M_2)$.
If $i < s$, then replace $i$ by $i+1$ and go back to step \textbf{is1}.
If $i = s$, then stop.
\end{description}
By using the method explained just below (\ref{eq:aug18_4}) we obtain a path $p$ from the final forms of the $M_1, M_2$.

\begin{theorem}[\cite{Takagi1}]\label{th:oct9_1}
The map $\Psi$ gives a bijection between $P_{L,M}$ (resp.~$P^+_{L,M}$)
and $\bigsqcup_{\lambda \vdash M} \{ \lambda \} \times \Omega(\lambda,L)$
(resp.~$\bigsqcup_{\lambda \vdash M} \{ \lambda \} \times \Omega^+(\lambda,L)$). 
\end{theorem}
\begin{theorem}[\cite{Takagi1}]\label{th:oct9_2}
Given $p \in P_{L,M}$ let $\Psi (p) = (\lambda, \boldsymbol{x})$.
Then $E_{\rm path}(p) = E_{\rm RC} (\lambda, \boldsymbol{x})$.
\end{theorem}

\section{Configuration sums over modified rigged configurations}\label{sec:5}

The $l$'s in this section should be interpreted as $2 \ell$'s in the previous section.

\begin{lemma}\label{lem:jun6_3}
For any positive integers $l,m,n,p$ the following relation holds
\begin{displaymath}
\sum_{x_1,\ldots,x_m,y} q^{x_1 + \cdots + x_m + y} = 
[\, p +  l (n+m) \,] 
\left[
\begin{array}{c}
p+m-1\\
m
\end{array}
\right],
\end{displaymath}
where the sum is taken over all integers $x_1,\ldots,x_m,y$ satisfying
\begin{eqnarray}
&& 0 \leq x_1 \leq \ldots \leq x_m \leq p + \min (x_1,l-1), \label{eq:july28_1}\\
&& n \max (0, x_m-p) \leq y \leq p-1+n \min (l,x_1). \label{eq:july28_2}
\end{eqnarray}
\end{lemma}
\noindent
\textit{Proof.}
Set $\lambda_1 = x_m - x_1, \lambda_2 = x_{m-1} - x_1, \ldots , \lambda_{m-1} = x_2 - x_1$ and $\lambda = (\lambda_1 ,\ldots , \lambda_{m-1})$.
Then
\begin{displaymath}
\sum_{x_1,\ldots,x_m,y} q^{x_1 + \cdots + x_m + y} = 
\sum_{x_1, y, \lambda} q^{m x_1 + y + |\lambda|},
\end{displaymath}
where the sum in RHS is over all integers $x_1, y$ satisfying
\begin{eqnarray*}
&& 0 \leq x_1 \leq p+l-\lambda_1-1, \\
&& n \max (0, x_1+\lambda_1-p) \leq y \leq p-1+n \min (l,x_1), 
\end{eqnarray*}
and over all partitions $\lambda$ with at most $m-1$ parts,
the largest part $\leq p$.
For any fixed $x_1, x_m$ the sum $\sum_{y: (\ref{eq:july28_2})} q^{y}$ amounts to
\begin{eqnarray}
\!\!\!\!\!\!\!\!\!\![p+n x_1], && (0 \leq x_1 \leq \min (l,p-\lambda_1) -1) 
\label{eq:jun6_1}\\
\!\!\!\!\!\!\!\!\!\!\left[ p+n l \right], && (l \leq x_1 \leq p-\lambda_1 -1) \\
\!\!\!\!\!\!\!\!\!\!q^{n x_1 -n (p - \lambda_1)} [p + n (p-\lambda_1)], && (p-\lambda_1 \leq x_1 \leq l-1) \\
\!\!\!\!\!\!\!\!\!\!\!\!\!\!\!\!\!\!\!\!q^{n x_1 -n (p - \lambda_1)} [p + n (p-\lambda_1 + l - x_1)]. && (\max(l,p-\lambda_1) \leq x_1 \leq p+l-\lambda_1-1)
\label{eq:jun6_2}
\end{eqnarray}
So if $l < p-\lambda_1$ the the sum $\sum_{x_1, y} q^{m x_1 + y}$ 
amounts to the sum of
\begin{eqnarray*}
&& \sum_{0 \leq x_1 \leq l-1} q^{m x_1} [p+n x_1] = \frac{1}{1-q} \left( 
\frac{[m l]}{[m]} - q^p \frac{[(m+n) l]}{[m+n]} \right), \\
&& \sum_{l \leq x_1 \leq p-\lambda_1-1} q^{m x_1} [p+n l]
= \frac{1}{1-q} \left(
(1-q^{p+l n}) \frac{[m (p-\lambda_1)] - [m l]}{[m]} \right), \\
&& \sum_{p-\lambda_1 \leq x_1 \leq p+l-\lambda_1-1} q^{(m + n) x_1 -n (p - \lambda_1)} [p + n (p-\lambda_1 + l - x_1)] \\
&& \qquad \qquad = \frac{1}{1-q} \left( 
q^{m (p-\lambda_1)} \frac{[(m+n) l]}{[m+n]} -
q^{p + l n + m (p-\lambda_1)}  \frac{[m l]}{[m]} \right),
\end{eqnarray*}
and if $l > p-\lambda_1$ it leads to sum of
\begin{eqnarray*}
&& \sum_{0 \leq x_1 \leq p-\lambda_1-1} q^{m x_1} [p+n x_1] = \frac{1}{1-q} \left( 
\frac{[m (p-\lambda_1)]}{[m]} - q^p \frac{[(m+n) (p-\lambda_1)]}{[m+n]} \right), \\
&& \sum_{p-\lambda_1 \leq x_1 \leq l-1} 
q^{(m + n) x_1 -n (p - \lambda_1)} [p + n (p-\lambda_1)] \\
&& \qquad \qquad = \frac{1}{1-q} \left(
(q^{-n (p-\lambda_1)}-q^{p}) \frac{[(m+n) l] - [(m+n) (p-\lambda_1)]}{[m+n]} \right), \\
&& \sum_{l \leq x_1 \leq p+l-\lambda_1-1} q^{(m + n) x_1 -n (p - \lambda_1)} [p + n (p-\lambda_1 + l - x_1)] \\
&& \qquad \qquad = \frac{1}{1-q} \left( 
q^{-n (p-\lambda_1) + (m+n)l} \frac{[(m+n)  (p-\lambda_1) ]}{[m+n]} -
q^{p + (m+n)l}  \frac{[m (p-\lambda_1)]}{[m]} \right).
\end{eqnarray*}
It is easy to see that in both cases one has
\begin{displaymath}
\sum_{x_1, y} q^{m x_1 + y} = 
\frac{1}{1-q} \left(
\frac{[l(m+n)]}{[m+n]} (q^{m(p-\lambda_1)} - q^p) +
\frac{[p+l(m+n)]}{[m]} (1 - q^{m(p-\lambda_1)}) \right).
\end{displaymath}
{}From this expression and by the following formulas we obtain the desired relation.
\begin{displaymath}
\sum_{\lambda} q^{|\lambda|} = 
\left[
\begin{array}{c}
p+m-1\\
m-1
\end{array}
\right],
\qquad
\sum_{\lambda} q^{|\lambda|+m(p-\lambda_1)} = q^p
\left[
\begin{array}{c}
p+m-1\\
m-1
\end{array}
\right].
\end{displaymath}
\hfill $\Box$

Let $l_1, \ldots ,l_s, m_1, \ldots , m_s$ be positive integers.
We impose $l_1 > \cdots >l_s$.
Given any integer $p_1 \geq 0$ we define $p_i \, (2 \leq i \leq s)$ by
$p_i = p_{i-1} + (m_1 + \cdots + m_{i-1}) (l_{i-1} - l_i)$.
Let $p_{s+1} = p_s + (m_1 + \cdots + m_s) l_s$

Let $\boldsymbol{x} = (x_{1,1}, \ldots , x_{1,m_1}, x_{2,1}, \ldots , x_{2,m_2}, \ldots ,
x_{s,1}, \ldots , x_{s,m_s}) \in \Z^{\mathcal N}$
where ${\mathcal N} = m_1 + \cdots + m_s$.
Then let 
\begin{displaymath}
\Omega_s = \left\{
\boldsymbol{x} \in \Z^{\mathcal N}
\vphantom{
\begin{array}{l}
0 \leq x_{i,1} \leq \cdots \leq x_{i,m_i} \leq p_i + \min (x_{i,1},l_i-1) \\
\mbox{for} \quad 1 \leq i \leq s, \\
x_{j,m_j} - p_j \leq x_{i,1}, \, x_{i,m_i} \leq p_i+l_i-l_j+x_{j,1}-1 \\
\mbox{for} \quad 1 \leq i < j \leq s
\end{array}
}
\right.
\left|
\begin{array}{l}
0 \leq x_{i,1} \leq \cdots \leq x_{i,m_i} \leq p_i + \min (x_{i,1},l_i-1) \\
\mbox{for} \quad 1 \leq i \leq s, \\
x_{j,m_j} - p_j \leq x_{i,1} \quad \mbox{for} \quad 1 \leq i < j \leq s, \\
x_{i,m_i} \leq p_i+l_i-l_j+x_{j,1}-1 \quad \mbox{for} \quad 1 \leq i < j \leq s
\end{array}
\right\} .
\end{displaymath}
If we set $l_i = 2 \ell_i$ and $p_{s+1} = L$, then $\Omega_s$ coincides with $\Omega^\circ (\lambda,L)$
in the previous section.
\begin{lemma}\label{lem:aug6_1}
The following relation holds:
\begin{equation}
\sum_{\boldsymbol{x} \in \Omega_s} q^{|\boldsymbol{x}|} =
\frac{[p_{s+1}]}{[m_1]}
\left[
\begin{array}{c}
p_1+m_1-1\\
m_1-1
\end{array}
\right]
\prod_{i=2}^{s}
\left[
\begin{array}{c}
p_i+m_i-1\\
m_i
\end{array}
\right],
\end{equation}
where $|\boldsymbol{x}| = \sum_{i=1}^s \sum_{k=1}^{m_i} x_{i,k}$.
\end{lemma}
\noindent
\textit{Proof.}
It is done by induction on $s$.
When $s=1$ the relation becomes
\begin{displaymath}
\sum_{\boldsymbol{x} \in \Omega_1} q^{x_1 + \cdots + x_m} =
\frac{[p+ml]}{[m]}
\left[
\begin{array}{c}
p+m-1\\
m-1
\end{array}
\right]
\end{displaymath}
where
\begin{displaymath}
\Omega_1 = \left\{
\boldsymbol{x} \in \Z^m
|
0 \leq x_{1} \leq \cdots \leq x_{m} \leq p + \min (x_{1},l-1)
\right\} .
\end{displaymath}
It is proved by using
\begin{displaymath}
\sum_{\boldsymbol{x} \in \Omega_1, x_1 < l-1} q^{x_1 + \cdots + x_m} =
(1 + q^m + \cdots + q^{(l-2)m})
\left[
\begin{array}{c}
p+m-1\\
m-1
\end{array}
\right] 
\end{displaymath}
and
\begin{displaymath}
\sum_{\boldsymbol{x} \in \Omega_1, x_1 \geq l-1} q^{x_1 + \cdots + x_m} =
q^{(l-1)m}
\left[
\begin{array}{c}
p+m\\
m
\end{array}
\right] .
\end{displaymath}

For $s > 1$ we assume the claim of Lemma but $s$ has been replaced by $s-1$.
More precisely, we set
\begin{equation}\label{eq:jun3_5}
\boldsymbol{x'} = (x_{1,1}, \ldots , x_{1,m_1}, x_{2,1}, \ldots , x_{2,m_2}, \ldots ,
x_{s-1,1}, \ldots , x_{s-1,m_{s-1}}) \in \Z^n
\end{equation}
where $n = m_1 + \cdots + m_{s-1}$ and let
\begin{displaymath}
\Omega_{s-1} = \left\{
\boldsymbol{x'} \in \Z^n
\vphantom{
\begin{array}{l}
0 \leq x_{i,1} \leq \cdots \leq x_{i,m_i} \leq p_i + \min (x_{i,1},l_i-1) \\
\mbox{for} \quad 1 \leq i \leq s-1, \\
x_{j,m_j} - p_j \leq x_{i,1}, \, x_{i,m_i} \leq p_i+l_i-l_j+x_{j,1}-1 \\
\mbox{for} \quad 1 \leq i < j \leq s-1
\end{array}
}
\right.
\left|
\begin{array}{l}
0 \leq x_{i,1} \leq \cdots \leq x_{i,m_i} \leq p_i + \min (x_{i,1},l_i-1) \\
\mbox{for} \quad 1 \leq i \leq s-1, \\
x_{j,m_j} - p_j \leq x_{i,1} \quad \mbox{for} \quad 1 \leq i < j \leq s-1,\\
x_{i,m_i} \leq p_i+l_i-l_j+x_{j,1}-1 \quad \mbox{for} \quad 1 \leq i < j \leq s-1
\end{array}
\right\}.
\end{displaymath}
Then the following relation is supposed to hold:
\begin{equation}\label{eq:jun3_6}
\sum_{\boldsymbol{x'} \in \Omega_{s-1}} q^{|\boldsymbol{x'}|} =
\frac{[p_{s-1}+n l_{s-1}]}{[m_1]}
\left[
\begin{array}{c}
p_1+m_1-1\\
m_1-1
\end{array}
\right]
\prod_{i=2}^{s-1}
\left[
\begin{array}{c}
p_i+m_i-1\\
m_i
\end{array}
\right].
\end{equation}
Note that $p_{s-1}+n l_{s-1 }= p_{s}+n l_{s}$.

We denote by $\pi$ the projection from $\Z^n \times \Z^{m_s}$ onto $\Z^n$.
Note that $\pi (\Omega_s) \subset \Z^n$ does not coincide with $\Omega_{s-1} \subset \Z^n$.
To see their differences, set $x_i = x_{s,i} \, (1 \leq i \leq m_s), m = m_s, l =l_s, p=p_s$, and
$\lambda_1 = x_m - x_1$.
Let $\Omega_s |_{x_1,\lambda_1}$ be the set of all $\boldsymbol{x} \in \Omega_s$ with fixed $x_1, \lambda_1$.
Then for any $\boldsymbol{x'} \in \pi (\Omega_s |_{x_1,\lambda_1})  \subset \Z^n$ the conditions
\begin{eqnarray}
&&x_1 + \lambda_1 - p \leq x_{i,1},\label{eq:jun3_1}\\
&&x_{i,m_i} \leq p_i + l_i - l+ x_1- 1,\label{eq:jun3_2}
\end{eqnarray}
for $1 \leq i \leq s-1$ are imposed,
in addition to
\begin{eqnarray}
&& 0 \leq x_{i,1} \leq \cdots \leq x_{i,m_i} \leq p_i + \min (x_{i,1},l_i-1) \quad \mbox{for} \quad 1 \leq i \leq s-1, \label{eq:jun3_3}\\
&& 
\begin{array}{l}
x_{j,m_j} - p_j \leq x_{i,1} \\
x_{i,m_i} \leq p_i+l_i-l_j+x_{j,1}-1
\end{array}
\quad \mbox{for} \quad 1 \leq i < j \leq s-1.
\label{eq:jun3_4}
\end{eqnarray}
Here we wrote the entries of 
$\boldsymbol{x'}$ as in (\ref{eq:jun3_5}).
Note that the latter conditions
(\ref{eq:jun3_3}), (\ref{eq:jun3_4}) are those for $\Omega_{s-1}$.
We claim that if the sum over $\boldsymbol{x'} \in \Omega_{s-1}$ in (\ref{eq:jun3_6}) is replaced by
the sum over {$\boldsymbol{x'} \in \pi ({\Omega}_{s}|_{x_1,\lambda_1})$},
then the only change thereby caused is
within the factor $[p_{s-1}+n l_{s-1}]$ in RHS.
More precisely, this factor is replaced by one of
(\ref{eq:jun6_1})-(\ref{eq:jun6_2})
depending on the conditions on $x_1$ and $\lambda_1$:
\begin{equation*}
\!\!\!\!\!\!\!\!\!\!\!\!\!\!\!\!\!\!\!\!\!\!\!\!\!\!\!\!\!\!\!\!
\sum_{\boldsymbol{x'} \in \pi ( \Omega_{s}|_{x_1,\lambda_1})} q^{|\boldsymbol{x'}|} =
Q \cdot 
\left\{
\begin{array}{ll}
[p+n x_1] & (0 \leq x_1 \leq \min (l,p-\lambda_1) -1) \\
\left[ p+n l \right] & (l \leq x_1 \leq p-\lambda_1-1) \\
q^{n (x_1 + \lambda_1 - p)} [p + n (p-\lambda_1)] & (p-\lambda_1 \leq x_1 \leq l-1) \\
q^{n (x_1 + \lambda_1 - p)} [p + n (p-\lambda_1 + l - x_1)] & (\max(l,p-\lambda_1) \leq x_1 \leq p+l-\lambda_1-1)
\end{array}
\right.
\end{equation*}
where
\begin{equation*}
 Q = \frac{1}{[m_1]}
\left[
\begin{array}{c}
p_1+m_1-1\\
m_1-1
\end{array}
\right]
\prod_{i=2}^{s-1}
\left[
\begin{array}{c}
p_i+m_i-1\\
m_i
\end{array}
\right].
\end{equation*}
Admitting this claim, the Lemma immediately follows as a result of
Lemma \ref{lem:jun6_3}:
\begin{eqnarray*}
&& \sum_{\boldsymbol{x} \in \Omega_{s}} q^{|\boldsymbol{x}|} =
\sum_{x_1,\ldots,x_m: (\ref{eq:july28_1})} \left( q^{x_1 + \cdots + x_m}
\sum_{\boldsymbol{x'} \in \pi(\Omega_{s}|_{x_1,\lambda_1})} q^{|\boldsymbol{x'}|} \right),\\
&&\sum_{\boldsymbol{x'} \in \pi(\Omega_{s}|_{x_1,\lambda_1})} q^{|\boldsymbol{x'}|} 
=Q \cdot \sum_{y: (\ref{eq:july28_2})} q^y,\\
&&\sum_{x_1,\ldots,x_m: (\ref{eq:july28_1}), y: (\ref{eq:july28_2})} q^{x_1 + \cdots + x_m + y} =
[p_s + l_s (m_1 + \cdots + m_s)]
\left[
\begin{array}{c}
p_s+m_s-1\\
m_s
\end{array}
\right].
\end{eqnarray*}

We suppose $0 \leq x_1 \leq \min (l,p-\lambda_1) -1$ and show how the factor $[p_{s-1}+n l_{s-1}]$ is replaced by $[p + n x_1]$.
The condition (\ref{eq:jun3_1}) is now satisfied as $x_1 + \lambda_1 - p \leq -1 < 0 \leq x_{i,1}$ by (\ref{eq:jun3_3}).
%
%
Based on the following arguments,
it is easy to see that $\pi ({\Omega}_{s}|_{x_1,\lambda_1})$ is
obtained from $\Omega_{s-1}$ by replacing $l_i$ with $l_i - l + x_1$ for all $1 \leq i \leq s-1$.
First of all, any simultaneous 
constant shift of $l_i$'s does not change $p_i$'s.
Hence it leaves the condition (\ref{eq:jun3_4}) unchanged, as well as all the factors in (\ref{eq:jun3_6}) 
except the $[p_{s-1}+n l_{s-1}]$. 
The condition (\ref{eq:jun3_2}) is achieved by the above replacement 
$l_i \rightarrow l_i - l + x_1$ in (\ref{eq:jun3_3}).
Therefore the 
only change of (\ref{eq:jun3_6}) caused by the replacement of $\Omega_{s-1}$
with $\pi ({\Omega}_{s}|_{x_1,\lambda_1})$ is given by replacing $[p_{s-1}+n l_{s-1}]$
with $[p_{s-1}+n (l_{s-1}-l+x_1)] = [p+n x_1]$.

Suppose $l \leq x_1 \leq p - \lambda_1 -1$.
The condition (\ref{eq:jun3_1}) is satisfied as above, and
(\ref{eq:jun3_2}) is now satisfied as $x_{i,m_i} \leq p_i + l_i - 1 \leq p_i + l_i - 1+ x_1- l$ by (\ref{eq:jun3_3}).
Hence no changes are caused by the replacement of $\Omega_{s-1}$
with $\pi ({\Omega}_{s}|_{x_1,\lambda_1})$. 
Note that $[p_{s-1}+n l_{s-1}] = [p + n l]$.

Suppose $p-\lambda_1 \leq x_1 \leq l-1$.
By the conditions (\ref{eq:jun3_1})-(\ref{eq:jun3_3}) we have
\begin{equation*}
x_1 + \lambda_1 - p \leq x_{i,1} \leq \cdots \leq x_{i,m_i} \leq p_i + \min (x_{i,1},l_i - l + x_1-1),
\end{equation*}
for $1 \leq i \leq s-1$.
By replacing $x_{i,j}$ with $x_{i,j} + x_1 + \lambda_1 - p$ this condition becomes
\begin{equation}\label{eq:oct13_1}
0 \leq x_{i,1} \leq \cdots \leq x_{i,m_i} \leq p_i + \min (x_{i,1},\tilde{l}_i -1),
\end{equation}
where $\tilde{l}_i = l_i - l - \lambda_1 + p$.
Note that the simultaneous constant shift $x_{i,j} \rightarrow x_{i,j} + x_1 + \lambda_1 - p$ for 
$\{ x_{i,j} \}_{1 \leq i \leq s-1, 1 \leq j \leq m_i}$ makes the change 
$q^{|\boldsymbol{x'}|} \rightarrow q^{|\boldsymbol{x'}| + n (x_1 + \lambda_1 - p)}$ and
leaves the condition (\ref{eq:jun3_4}) unchanged.
Hence
\begin{equation*}
\sum_{\boldsymbol{x'} \in \pi(\Omega_{s}|_{x_1,\lambda_1})} q^{|\boldsymbol{x'}|} =
q^{ n (x_1 + \lambda_1 - p)} \sum_{\boldsymbol{x'}: (\ref{eq:jun3_4}),(\ref{eq:oct13_1})}
q^{|\boldsymbol{x'}|}.
\end{equation*}
By the replacement 
$l_i \rightarrow \tilde{l}_i$ the condition (\ref{eq:jun3_3}) becomes (\ref{eq:oct13_1}), and
the condition (\ref{eq:jun3_4}) is left unchanged.
Hence
\begin{equation*}
\sum_{\boldsymbol{x'}: (\ref{eq:jun3_4}),(\ref{eq:oct13_1})}
q^{|\boldsymbol{x'}|} = Q \cdot [p_{s-1} + n \tilde{l}_{s-1}] = Q \cdot [p + n (p-\lambda_1)].
\end{equation*}

Suppose $\max(l,p-\lambda_1) \leq x_1 \leq p+l-\lambda_1-1$.
The condition (\ref{eq:jun3_2}) is now satisfied as $x_{i,m_i} \leq p_i + l_i - 1 \leq p_i + l_i - 1 + x_1 - l$ by (\ref{eq:jun3_3}).
By the conditions (\ref{eq:jun3_1}), (\ref{eq:jun3_3}) we have
\begin{equation*}
x_1 + \lambda_1 - p \leq x_{i,1} \leq \cdots \leq x_{i,m_i} \leq p_i + \min (x_{i,1},l_i -1),
\end{equation*}
for $1 \leq i \leq s-1$.
By replacing $x_{i,j}$ with $x_{i,j} + x_1 + \lambda_1 - p$ this condition becomes
\begin{equation}\label{eq:oct13_2}
0 \leq x_{i,1} \leq \cdots \leq x_{i,m_i} \leq p_i + \min (x_{i,1},\hat{l}_i -1),
\end{equation}
where $\hat{l}_i = l_i - x_1 - \lambda_1 + p$.
By the same argument in the previous paragraph, we have
\begin{eqnarray*}
&& \sum_{\boldsymbol{x'} \in \pi(\Omega_{s}|_{x_1,\lambda_1})} q^{|\boldsymbol{x'}|} =
q^{ n (x_1 + \lambda_1 - p)} \sum_{\boldsymbol{x'}: (\ref{eq:jun3_4}),(\ref{eq:oct13_2})}
q^{|\boldsymbol{x'}|}, \\
&& \sum_{\boldsymbol{x'}: (\ref{eq:jun3_4}),(\ref{eq:oct13_2})}
q^{|\boldsymbol{x'}|} = Q \cdot [p_{s-1} + n \hat{l}_{s-1}] = Q \cdot [p + n (p-\lambda_1 + l - x_1)].
\end{eqnarray*}
\hfill $\Box$

\noindent
{\it Acknowledgement}.
This work is supported by Grants-in-Aid for 
Scientific Research No.~21540209 and No.~25540241
from JSPS.

\vspace{0.5cm}


\begin{thebibliography}{99}

\bibitem{PC}Kashiwara M and Miwa T (eds) 2000
\textit{Physical Combinatorics}, 
Prog. in Math. {\bf 191} (Boston: Birkh\"auser) 

\bibitem{PC2}Kuniba A and Okado M (eds) 2007
\textit{Combinatorial aspect of integrable systems},
Math. Soc. Jpn.  Memoirs {\bf 17} (Tokyo: MSJ)

\bibitem{Ba}Baxter R J 2007
\textit{Exactly solved models in statistical mechanics} (New York: Dover)

\bibitem{KKR}
Kerov S V, Kirillov A N and Reshetikhin N Yu 1988
Combinatorics, the Bethe ansatz and representations 
of the symmetric group. 
J. Soviet Math. {\bf 41} 916--924 

\bibitem{KR} Kirillov A N and Reshetikhin N Yu 1988
The Bethe ansatz and the combinatorics of Young tableaux,
J. Soviet Math. {\bf 41} 925--955

\bibitem{Be} Bethe H A 1931
Zur Theorie der Metalle. I.~Eigenwerte 
und Eigenfunktionen der linearen Atomkettee,
{\it Z. Phys.} {\bf 71} 205--226

\bibitem{TS} Takahashi D, Satsuma J 1990
A soliton cellular automaton,
J. Phys. Soc. Japan {\bf 59}  3514--3519

\bibitem{KOTY}Kuniba A, Okado M, Takagi T and Yamada Y 2003
Vertex operators and canonical partition function of box-ball system
(in Japanese), RIMS K$\hat{\rm o}$ky$\hat{\rm u}$roku, 
{\bf 1302} 91--107

\bibitem{Takagi1} Takagi T 2005
{Inverse scattering method for a soliton cellular automaton}
Nucl. Phys. B {\bf 707} 577--601


\bibitem{YYT}
Yoshihara D, Yura F and Tokihiro T 2003
 Fundamental cycle of a periodic box-ball system,
J. Phys. A: Math. Gen. {\bf 36}  99--121

\bibitem{KTT} Kuniba A Takagi T and Takenouchi A 2006 
Bethe ansatz and inverse scattering transform 
in a periodic box-ball system
Nucl. Phys. B {\bf 747} [PM] 354--397

\bibitem{KN} Kuniba A and Nakanishi T 2000
The Bethe equation at $q=0$, the M\"obius inversion formula, 
and weight multiplicities: I. 
The $sl(2)$ case, Prog. in Math. {\bf 191} 185--216

\bibitem{A}
Andrews G E 1984
\textit{The Theory of Partitions}
(Cambridge: Cambridge University Press.)

\bibitem{St}
Stanley R P 1999
\textit{Enumerative combinatorics} vol.1
(Cambridge: Cambridge University Press.)

\bibitem{KS} 
Kirillov A N and Sakamoto R 2009
Relationships between two approaches: 
rigged configurations and 10-eliminations,
Lett.  in Math. Phys {\bf 89}  51--65 
 
\bibitem{MIT}
Mada J Idzumi M and Tokihiro T 2006
On the initial value problem of a periodic box-ball system
J. Phys. A: Math. Gen. {\bf 39} L617


\end{thebibliography}
\end{document}